# Economic Consequences of Online Tracking Restrictions:

# Evidence from Cookies

Klaus M. Miller

Assistant Professor of Quantitative Marketing

Department of Marketing, HEC Paris

1 Rue de la Libération, 78350 Jouy-en-Josas

Phone: +33 1 39 67 70 88

Email: millerk@hec.fr

Bernd Skiera*

Chaired Professor of Electronic Commerce

Faculty of Economics and Business, Goethe University Frankfurt

Theodor-W.-Adorno-Platz 4, 60323 Frankfurt am Main

Phone: +49 69 798 34649

Email: skiera@wiwi.uni-frankfurt.de

* We acknowledge financial support of the NET Institute at the NYU Stern School of Business, the Hi!PARIS Center in Data Analytics & AI for Science, Technology, Business & Society, and from the European Research Council (ERC) under the European Union's Horizon 2020 Research and Innovation Program (Grant Agreement No 833714).

# Economic Consequences of Online Tracking Restrictions:

# Evidence from Cookies

In recent years, European regulators have debated restricting the time an online tracker can track a user to protect consumer privacy better. Despite the significance of these debates, there has been a noticeable absence of any comprehensive cost-benefit analysis. This article fills this gap on the cost side by suggesting an approach to estimate the economic consequences of lifetime restrictions on cookies for publishers. The empirical study on cookies of 54,127 users who received ~128 million ad impressions over ~2.5 years yields an average cookie lifetime of 279 days, with an average value of €2.52 per cookie. Only ~13% of all cookies increase their daily value over time, but their average value is about four times larger than the average value of all cookies. Restricting cookies' lifetime to one year (two years) decreases their lifetime value by ~25% (~19%), which represents a decrease in the value of all cookies of ~9% (~5%). In light of the €10.60 billion cookie-based display ad revenue in Europe, such restrictions would endanger €904 million (€576 million) annually, equivalent to €2.08 (€1.33) per EU internet user. The article discusses these results' marketing strategy challenges and opportunities for advertisers and publishers.

# 1. Introduction

Marketers are still increasing their digital marketing by allocating 53.8% of their budgets to digital marketing activities. They plan to increase spending by another 5.7% in 2024 (March 2023 edition of The CMO Survey). One reason for these increased investments is the marketers' capability to track users along their customer journey, as (McKee 2021) outlines: "Being able to accurately measure the reach and calculate the ROI of digital campaigns is hugely attractive to CMOs. Not only does this give them the information they need to optimize and improve the effectiveness of their digital spending, but it also arms them with quantifiable facts and figures to take to their boards and justify their positions."

Such tracking is possible because of the digital environment and the availability of tracking technologies such as (digital) cookies, digital fingerprinting, session IDs, and logins. 85% of marketers say their marketing activities are slightly to completely reliant on third-party cookies (HubSpot 2023). Nevertheless, marketers will need to reconcile this strategy with looming changes in privacy regulations (Acquisti 2023; Johnson 2023).

Tracking allows firms to acquire more profound insights into user behavior, thereby improving the user experience by implementing strategies for personalizing content. Nevertheless, the gathered data also serves in deploying other marketing tactics such as attribution modeling and, notably, targeted advertising - frequency capping included. Better targeting of ads comes in many forms, but retargeting is undoubtedly one of the most remarkable (Lambrecht and Tucker 2013; Bleier and Eisenbeiss 2015). A typical setting for retargeting is an online shop where a user puts a product into a shopping basket but does not purchase it. The online shop can now inform a retargeting provider, such as Criteo, about this behavior. The retargeting provider then puts up ads for the online shop and the abandoned product on many other websites so that the user will observe an ad about the specific product on another website (e.g., an online newspaper), even if this website is unrelated to the online shop.

Unsurprisingly, many users are increasingly concerned about their privacy in such cases (Bleier, Goldfarb, and Tucker 2020; Beke et al. 2022). They often feel surveilled; some even find it "creepy" that a website can show them ads related to their behavior elsewhere. They may not know that (1) a tracking technology—here, cookies—enabled the online shop to track their behavior; (2) cookie matching informed the retargeting provider about the online shop's goal to target the user with ads that relate to the abandoned products; and (3) the website that showed the ad essentially had no information about what the user did in the online shop.

Policy makers are concerned and want to increase privacy. Many privacy regulations, including the European General Data Protection Regulation (GDPR) and China's Personal Information Protection Law (PIPL), emphasize obtaining a user's consent to tracking (Jin and Skiera 2022). However, they are less focused on how long firms can store data after obtaining a user's consent, even though the duration of data storage is key for privacy protection and the security of a user's data. Therefore, regulators, such as in the EU with the upcoming ePrivacy Regulation (Council of the European Union 2021), have discussed limiting the use of tracking technologies and data exchange between firms, among them restricting the lifetime of information collected by tracking technologies such as cookies.

Such a restriction implements the users' (human) right to be forgotten, which is implemented in several privacy laws such as GDPR and reflects that firms (or other entities) must delete certain user data after a specific period. However, evaluating the cost of such a lifetime limitation is challenging for at least two reasons. First, little knowledge exists about how long cookies even live. Many users claim they delete cookies regularly; however, the privacy paradox shows that users' stated and actual behavior can differ substantially (Gross and Acquisti 2005; Skiera et al. 2022). Second, we know little about the size of the economic consequences for publishers that would result from such a lifetime restriction. A loss is likely to occur as the information collected via cookies enables the advertiser to target ads better. Nevertheless, it is unclear how significant the loss is.

This lack of knowledge is unfortunate because online advertising revenue ($209.7 billion (€86 billion) in 2022 in the United States (Europe) alone; Interactive Advertising Bureau 2023a; 2023b) finances the content available to users, often free of charge or at least at a relatively low price. As a result, policy makers have no way of telling whether any benefits for consumer privacy that come with restrictions on tracking technologies are outweighed by adverse effects on publishers' profits—which policy makers have to trade off carefully.

Herein, we aim to fill this void by suggesting and applying an approach for estimating the economic consequences of restricting cookie lifetimes for publishers. Specifically, we compute the lifetime value of cookies (LVC) and then determine the economic consequences (in our case, a loss) of cookie lifetime restrictions for publishers. We test which cookies live longer than a given lifetime restriction and grow in value over time. We find, for example, that only ~13% of all cookies increase their daily value over time, but their average value is about four times larger than the average value of all cookies. We then derive the economic loss that would result from restricting cookie lifetime—for example, to 12 months as the European Union has proposed (Article 29 Data Protection Working Party 2010; European Union 2017a; 2017b, Council of the European Union 2021) and the Italian DPA has already implemented or to 24 months, as the Spanish DPA and some industry stakeholders such as Facebook (Cook 2017) have implemented.

We analyze data covering about 2.5 years collected by a large European ad exchange that serves as a programmatic marketplace to facilitate automated buying and selling of online display, mobile, and video advertising inventory. This panel enables us to use within-cookie price variations to trace the development of the value of a cookie over time, which no previous researchers have achieved.

Our results contribute to ongoing discussions in the digital advertising industry on the consequences of increasing online tracking restrictions to protect consumer privacy. They provide insights for the industry, users, and policy makers, such as determining how much users should pay for being tracked less intensively. Such knowledge could serve as a basis for regulatory decisions, industry self-regulation efforts (e.g., programs offering tracking-free digital subscriptions used by European publishers such as the Standard in Austria, Spiegel in Germany, or the premium EU subscription provided by The Washington Post ), and privacy taxation. Furthermore, Chief Marketing Officers (CMOs) gain insights into the value of user data and targeting strategies. For example, restrictions on collecting user data would force firms to move away from behavioral targeting and focus on other targeting opportunities, e.g., contextual or geo-targeting. So, our results add to the literature on quantifying the value of the information generated in digital markets and determining the price of privacy.

## 2. Overview of Online Tracking Technologies

In basic terms, user tracking identifies a user online via a unique identifier and records the user's behavior over time (Skiera et al. 2022). Tracking is easy if the user is willing to log in to a particular website because that login identifies the user on different browsers and devices. However, users are usually not ready to identify themselves via logins unless they benefit significantly from doing so. Social networks like Facebook provide such a benefit, but most other firms do not. As a result, they cannot rely on login data and must instead use other tracking technologies to identify users.

Mayer and Mitchell (2012) distinguish between stateless and stateful online tracking technologies. Stateless tracking does not store information but identifies the user by the (almost unique) configuration of the user's device (Besson et al. 2013). Stateful tracking stores the information required to identify a user on the user's device (also called the user client), usually a desktop computer, tablet, or smartphone (Sanchez-Rola et al. 2016). Note that the data associated with a user's identifier, such as their consumer profile, can be stored on the user's device or the firm's server tracking the user, the latter of which is currently common.

*2.1 Stateless Online Tracking*

Stateless online tracking mainly refers to digital fingerprinting. It exploits the fact that the configuration of the user's device has many attributes (e.g., CPU type, computer clock skew, display settings, scripts, browser, operating system information, IP address, and language settings; Mayer and Mitchell 2012). As a result, the specific combination of attribute values on a particular device is (almost) unique. Thus, observing the same configuration at multiple points makes it very likely that the configuration belongs to only one user. Behavioral biometric features—namely, dynamics that occur when typing, moving, clicking the mouse, or touching a touch screen—can provide further information to improve digital fingerprinting and, hence, user identification (Pugliese 2015). However, a disadvantage of this tracking technology is that a firm can no longer track a user if changes in the configuration of the user's device occur.

*2.2 Stateful Online Tracking*

Stateful online tracking techniques mainly refer to cookies. In simplified terms, a cookie is a small piece of data sent from a server (i.e., a website) to a browser and stored on the user's device (Cristal 2014). The most common are HTTP cookies (herein referred to simply as "cookies"), the focus of this study. They are stored in the user's browser, so they only identify the user if they continue using the same browser. Users can control such storage because all significant browsers enable users to prevent or delete cookies. Typically, a website is not incentivized to shorten a cookie's lifetime.

Cookies come in two main types: first-party cookies are installed by the website the user is visiting, whereas third-party cookies are installed by a server that does not belong to the website the user is visiting (e.g., a third-party ad server). First-party cookies only allow a firm to track a user on the website that initially set the cookie, whereas third-party cookies allow a firm to track a user across different websites. A cookie usually contains a unique number called "cookie-ID" that identifies the user (e.g., "179'032'342'526'846'362" in our data). Each cookie also has an expiration date that dictates when the browser will automatically delete the cookie. Table W2.1 (in Web Appendix W2) shows the maximum cookie lifetimes from a single visit to selected domains, which vary between approximately 1 and 20 years. Table W2.1 also outlines that a website often sets more than one cookie.

A cookie is typically associated with a single user. Sometimes, however, multiple users (e.g., from the same household) may share one browser on one device, such that one cookie captures the activities of multiple users. More frequently, however, a single user operates on multiple devices (e.g., desktop, tablet, smartphone) and even multiple browsers on one device (Budak et al. 2016; Yan, Miller, and Skiera 2022). In such a case, multiple cookies store the activities of a single user. Unfortunately, those cookies are often not connected, which represents a shortcoming of this tracking technology because it leads to identity fragmentation (Lin and Misra 2022; Min, Hoban, and Arora 2023).

In our study, we focus on cookies for three reasons. First, despite the availability of other online tracking technologies such as digital fingerprinting or advertising identifiers and plans to phase out third-party cookies, cookies are still frequently used to track a user on a single device on a specific website or across several websites.

Second, even if cookies are being (partially) replaced, the need for online tracking will prevail as the online advertising industry still wants to track, profile, and target online users to increase the efficiency of online advertising. As cookies work similarly to other online tracking technologies, such as digital fingerprinting, our results will likely generalize to these other tracking technologies that track users similarly to cookies.

Third, privacy regulation, e.g., the European Union's General Data Protection Regulation (GDPR; see Miller, Schmitt, and Skiera 2023) or the upcoming ePrivacy Regulation, generally regulate online tracking technologies and are agnostic towards a specific tracking technology, such as cookies. The initiatives to restrict an online tracking technology's lifetime (see Section 3.3) could therefore apply to cookies (as we study in this research) but also to other online tracking technologies.

## 3. Initiatives to Restrict Online Tracking

*3.1 Firms' Usage of Online Tracking*

Firms primarily use online trackers such as cookies or digital fingerprints to implement two marketing strategies: First, trackers enable publishers to personalize content to benefit users' experience. For example, websites use trackers to remember users' preferences, set up personalized content, and help users complete tasks without reentering information when revisiting a website (Cristal 2014; Tam and Ho 2006). Usually, a first-party tracker is sufficient to accomplish this personalization (Estrada-Jiménez et al. 2017).

Second, online trackers enable advertisers to target users with ads better, rendering their ads more effective (Goldfarb and Tucker 2011b). Particular forms of online advertising would probably not even exist without trackers. For example, it would be challenging to attribute purchases to affiliates without trackers in affiliate marketing. Better advertising can also benefit users, who receive more relevant ads to help them make better purchase decisions (for a summary of studies, see Boerman, Kruikemeier, and Borgesisus 2017; and a recent critique that targeted ads may not always be welfare-enhancing for users from Mustri, Adjerid, and Acquisti 2023). Summers, Smith, and Reczek (2016) have shown that targeted ads can even change how users think about themselves. Still, the decrease in privacy that comes with better-targeted ads may offset those benefits.

Advertisers usually use third-party trackers for ad targeting, allowing them to track users across websites and apps. Users may perceive such cross-site and cross-app tracking as infringing their privacy. Thus, several initiatives exist to restrict third-party trackers (cookies) but not necessarily first-party trackers (cookies). Subsequently, we will outline major privacy initiatives by policy makers and self-regulatory initiatives of the online advertising industry to restrict the usage of third-party tracking, especially cookies, and replace this form of tracking with less privacy-invasive ways of tracking. We note that although specific forms of tracking, such as third-party cookies, may lose importance, publishers will continue to benefit from their strategy of tracking users because it seems to help them better monetize their content online, as advertisers will benefit from a lower ad wastage.

*3.2 Initiatives to Restrict Online Tracking*

As tracking technologies such as cookies decrease user privacy (Awad and Krishnan 2006), policy makers and the online advertising industry have begun restricting online tracking technologies to protect users' privacy (see Figure 1).

< INSERT FIGURE 1 ABOUT HERE >

In Europe, online tracking is mainly governed by the ePrivacy Directive (ePD) and the General Data Protection Regulation (GDPR). The GDPR, for example, only allows the processing of personal data of internet users if one of the following applies: (a) the user has given consent to the processing of his or her personal data, (b) the processing is necessary for the performance of a contract, or (c) processing is in the legitimate interests pursued by the firm (other legal bases exist but are not applicable in the online setting).

Concerning online tracking, some European DPAs clarified that individual opt-in consent is the most appropriate legal basis for websites to process a user's personal data (Article 29 Data Protection Working Party 2012; Data Protection Commission 2020). Yet, the EU DPAs have signaled some openness to online trackers that exclusively use the website's own data (e.g., using first-party cookies) but remain concerned about online trackers that combine user data across websites (e.g., using third-party cookies). So, websites may deviate from using opt-in consent as a legal basis for online tracking and instead rely on another legal basis, such as legitimate interest. In contrast to GDPR, the California

Consumer Privacy Act (CCPA) defaults a user into online tracking but allows a user to opt-out if she does not want to be tracked (Jin and Skiera 2022).

In addition to initiatives of policy makers, many browsers increasingly protect online consumer privacy by restricting third-party online tracking as part of self-regulatory efforts. As early as 2013, Mozilla Firefox and Apple Safari started blocking third-party cookies from advertisers. In 2019, Apple began blocking all third-party tracking by launching its second version of Intelligent Tracking Prevention (ITP) in its Safari browser. Microsoft Edge has blocked third-party cookies since July 2020. In 2021, Apple followed suit and began preventing third-party tracking on mobile devices by introducing the App Tracking Transparency (ATT) framework.

Similarly, Google has set a 2024 deadline for preventing cross-site and cross-app tracking and announced eliminating third-party cookies in its Chrome browser (Google 2022a). Since June 2022, Mozilla has offered Total Cookie Protection (TCP), which confines third-party cookies to the site where they were created by treating them as first-party cookies. So, the trackers can see a user's behavior on the website the user visited but not the behavior on other websites.

Although the online advertising industry is moving toward a world without (third-party) cookies, the industry is also working on privacy-sensitive replacements for third-party cookies as the need for targeting ads remains, as does the need to monetize content. For example, as a substitute for device-level user tracking and advertising attribution, Apple created SKAdNetwork (SKAN) as a measurement solution to provide advertisers with coarse, time-delayed feedback on their post-ad click conversion events. Chrome's Privacy Sandbox separates the functions performed by third-party trackers into separate replacement technologies for ad targeting (FloC, FLEDGE), ad measurement (Aggregate & Conversion Measurement APIs), and other uses like fraud prevention (Trust Token API). Recently, Google introduced its Topics API for interest-based advertising, which replaced Google's earlier FloC proposal (Google 2022b). Such an approach may improve the outcome of online advertising (i.e., users are not individually targeted), but not necessarily the process of online advertising (i.e., users are still individually tracked on their devices). Raising questions of whether consumers perceive their privacy violated when their data is still being tracked, albeit anonymously (Jerath and Miller 2023).

In addition, the online advertising industry is working on strategies to avoid third-party cookies. They increase their use of first-party cookies and other identifiers such as a user's email address or login data. They also work on privacy-preserving strategies for sharing data among companies (Schneider et al. 2017; Kakatkar and Spann 2019). For example, European publishers created netID, a foundation established by an alliance of publishers to increase user tracking across the associate publishers based on the user's login (Skiera et al. 2022). The basic premise of netID is that it provides a user with a single (netID) account to access different publishers and manage all permission decisions. This centralization will likely reduce users' decision costs when providing and managing permission for data processing. As netID only relies on first-party tracking, it does not fall under the above-outlined restrictions of third-party tracking. However, it recreated the functionality of third-party tracking using first-party tracking across multiple publisher websites.

We can observe similar developments in mobile advertising, where Apple's strategy with its App Tracking and Transparency (ATT) intends to hamper third-party interactions in mobile applications. Yet, some advertisers circumvent this practice and replace third-party trackers using a fingerprinting-derived identifier (Kollnig et al. 2022).

The online publishing industry is also innovating by offering its users the opportunity to pay to avoid being tracked. For example, in May 2018, an Austrian publisher, the news website "Der Standard", introduced a notification banner, referred to as a pay-or-consent wall, that appears for first-time users of their website. It offers the user two options: (i) pay for not being tracked and not seeing advertisements or (ii) consent to being tracked, involving the processing of their personal data for third-party advertising and seeing ads. If the user refuses to select one of the two options, accessing the publisher's content is impossible, meaning the user has to take the third option and leave the website.

Finally, another way to restrict online tracking is to introduce a Digital Service Tax. Most DSTs became effective in 2020 and most significantly affect business models that derive a high value from consumer interaction (e.g., social networks and search engines). Their tax rates vary from 2.0% to 7.5%. All DSTs apply only to those firms whose global (i.e., worldwide) and local (i.e., country-specific) revenue surpasses a substantial threshold. While all DSTs build upon revenues, they differ regarding the taxed revenue type, precisely advertising revenue. For example, the tax bases in Austria, Italy, Turkey, and the United Kingdom are online advertising revenues. In contrast, the tax base in France is only targeted online advertising revenue.

*3.3 Initiatives to Restrict Tracking Technologies' Lifetime*

Currently, many privacy regulations emphasize obtaining a user's consent for tracking (e.g., the GDPR) and focus less on how long to store data after a user's consent has been obtained. However, the duration of data storage is vital for privacy protection and a user's data security. Therefore, EU regulators have discussed limiting the use of tracking technologies and data exchange among firms, among them the restriction of the lifetime of information collected by tracking technologies such as cookies (i.e., the right to be forgotten; Agencia Española de Protección de Datos 2020; Article 29 Data Protection Working Party 2010; Commission Nationale de l'Informatique et des Libertés 2020; Garante per la Protezione dei Dati Personali 2015; Voisin et al. 2021). For example, the Council of the European Union (2021) suggested a 12-month restriction of a tracker's lifetime in its draft for the ePrivacy Regulation. Such lifetime restrictions are already legally binding in some EU member states, such as France (6 months), Italy (12 months), and Spain (24 months). Other EU member states, such as Germany, for example, only advocate relatively shorter lifespans but do not specify precisely how long the useful life of a tracking technology should be.

Figure 2 summarizes these planned and partially implemented initiatives of EU member states. Web Appendix W1 describes these initiatives in more detail. Some firms have already implemented a rather loose cookie lifetime restriction (i.e., two years in the case of Facebook or 18 months in the case of Google), with the notable exception of Tradelab (2019), who adopted a rather strict cookie lifetime restriction of 13 months. This voluntary industry self-regulation may prevent too strict regulation by data protection agencies and regulators. Also, big techs such as Facebook and Google may be better able to substitute an information loss from one tracking technology (e.g., cookies) with other tracking technologies such as logins, for which lifetime restrictions are currently not discussed.

We note, however, that these restrictions could, at least theoretically, apply to all tracking technologies. So, they would equally affect cookies, digital fingerprints, logins, and other tracking technologies. The different implemented and planned cookie lifetime restrictions further require deleting the cookie after a specific time. That is, the activity or inactivity of a cookie does not play a role in the cookie lifetime restriction to kick-in.

< INSERT FIGURE 2 ABOUT HERE >

## 4. Knowledge on Cookies

*4.1 Knowledge on the Lifetime of Cookies*

Table 1 provides an overview of previous research on cookies. Primarily, it focuses on the value of cookies. Prior knowledge on cookie lifetime is limited because existing studies only report cookie age, corresponding to a cross-sectional measure of cookie lifetime at the specific time of data collection. For example, in assessing the impact of cookie deletion on website and ad metrics, Abraham, Meierhoefer, and Lispman (2007) find that 31% of U.S. users delete their cookies within one month. Beales and Eisenach (2014) report a mean cookie age of 1.5 months, and Johnson, Lewis, and Nubbemeyer (2017) find a median cookie age of 3.5 months. Such knowledge is valuable because it serves as a lower bound for the cookie's lifetime, though not as a proxy for the lifetime itself. It also shows that the cookie age is low, with a maximum of 3.5 months.

< INSERT TABLE 1 ABOUT HERE >

*4.2 Knowledge on the Value of Cookies*

Many discussions on the value of cookies include debates on the economic loss to publishers that a complete ban on cookies would entail. McKinsey (2010) estimates a monthly willingness to pay to avoid advertising intrusion as high as €10 per household. Budak et al. (2016) estimate that $2 per month would be enough to offset all digital ad revenues for content publishers. Amazon (Dastin 2019) offered users $10 in exchange for tracking them all over the web. In 2020, German publisher Spiegel introduced its "Pure" digital subscription, which charges €5 (≈$5.38) per month for an ad- and tracking-free online browsing experience. The Washington Post offers a similar paid and ad-free offer to its EU users for €4 per month. Finally, Deighton and Kornfeld (2020) estimate that if all online tracking were to end, absent a mitigating technology, there would be a shift of between $32 billion and $39 billion of advertising and ecosystem revenue away from the open web toward walled gardens such as Meta, Google or Apple.

However, current insights on the value of a cookie diverge greatly. Beales's (2010) early survey of advertising networks suggests that tracking commands a price premium: average CPMs ("cost per 1,000 impressions") were $1.98 for untargeted "run of network" ads, $4.12 for behavioral targeting, and $3.07 for retargeting. Beales and Eisenach (2014) show a positive impact of cookies on advertisers' willingness to pay for ads; cookies result in a premium of $.22 to $.34 for 1,000 impressions (reflected in the respective increase of CPM). Further, they quantify the positive impact of cookies' age on advertisers' willingness to pay between $.02 and $.15 (CPM) per month as users

with older cookies accumulate more information. Goldfarb and Tucker (2011a) find that advertising was approximately 65% less effective after a European cookie restriction policy.

Johnson's (2013) results indicate that a complete tracking ban (i.e., no cookies are allowed) would lead to a drop in publisher revenues by 38.5% and a 45.5% drop in advertiser surplus. Aziz and Telang (2016) find that more intrusive information makes ad targeting better but at a decreasing rate. Using an industry field experiment, Ravichandran and Korula (2019) find a decrease in publisher revenue by 52% in the absence of third-party cookies. Rafieian and Yoganarasimhan's (2021) assessment of the value of mobile targeting shows that behavioral targeting leads to a 12.1% improvement in advertisers' targeting ability, as measured by the relative information gain of behavioral over nonbehavioral targeting. Johnson, Shriver, and Du (2020) estimate the total lost expenditure on behavioral targeting to be $8.58 per opt-out user for the U.S. desktop display industry in 2015, corresponding to a 52% loss in ad prices as paid by advertisers without a cookie, while Wang, Jiang, and Yang (2023) only find a 6% reduction in advertiser bid prices and the resulting revenues per click.

Similarly, Laub, Miller, and Skiera (2023) find an 18% decrease in ad prices paid to publishers without a cookie. In contrast, Marotta, Abishek, and Acquisti (2019) estimate an ad price loss of only 8% to publishers. Overall, the studies mentioned above show that cookies generate value, but there is considerable heterogeneity between studies on how to measure the value of a cookie and how large this value should be; losses range from 6% to 65%.

*4.3 Knowledge on the Development of the Value of Cookies*

Knowledge about the development of a cookie's value over time is sparse. Some evidence in the literature indicates that a cookie's value increases over time (e.g., Beales and Eisenach 2014; Casale 2015). The core argument for an increase in value is that advertisers collect more information on a user's profile, and this information enables better targeting. Empirical research in the domain of targeting has primarily focused on elucidating the relationship between consumer targeting — as reflected in reliance on ad targeting (versus no targeting) — and the profits of publishers and advertisers. These studies show that targeting is more profitable for publishers and advertisers than no targeting.

More specifically, these studies find positive effects of targeting on a variety of outcome measures, including the following: click-through rates (Yan et al. 2009), which depend on timing and placement factors (Bleier and Eisenbeiss 2015); consumers' purchase intentions (van Doorn and Hoekstra 2013); purchase probabilities (Manchanda et al. 2006); consumers' progression through the purchase funnel (Hoban and Bucklin 2015); ad prices and ad revenue (Beales 2010), ad revenue per impression (Ada, Abou Nabout, and Feit 2022); and advertising profitability (Lewis and Reiley 2014). These varying outcomes also explain differences in the value of users for publishers.

In contrast, prior theoretical studies on consumer targeting have identified situations in which advertisers or publishers have incentives to prevent the targeting of ads based on a large number of attributes (Badanidiyuru, Bhawalkar, and Xu 2018). Levin and Milgrom (2010) explain the intuition behind this remarkable finding. If competition for a particular ad exposure is low—e.g. when very few advertisers wish to display an ad to a specific consumer—the publisher suffers from lower prices. The likelihood of such a "thin market" increases with the number of attribute levels in consumer profiles because some attributes can indicate that a consumer is unattractive to an advertiser. For example, knowing that someone is male keeps him in the target group for a fashion store for young people, but knowing that the person is male and 70 years old might exclude him.

Shedding further light on these ideas, a study by Board (2009) shows theoretically that, in a second-price sealed-bid auction, the number of attribute levels in a consumer profile is always positively related to the sum of the profits gained by the publisher and advertisers. However, the number of attributes does not straightforwardly affect how this profit splits between advertisers and the publisher. In some cases, as outlined above, much information about consumers can decrease prices but increase prices in other cases.

In sum, the theoretical studies cited above suggest that including more attributes in consumer profiles can often, but does not always, lead to higher prices. They also outline reasons why the effects could differ across users. For example, suppose a user belongs to a small group of users, i.e., a thin market. Little competition might exist for targeting this user, so the value of the user for the publisher is low. In contrast, suppose a user belongs to a much larger group of users. Competition for targeting these users might be much higher, as does the value of the user for the publisher.

The literature on retargeting (Bleier and Eisenbeiss 2015) and informal discussions with industry practitioners suggest that a user may be only in the market for a specific product for a short time (e.g., renting a car for a weekend trip to San Francisco but otherwise not interested in renting a car in San Francisco). In such a case, the value of a cookie would be higher at the beginning of the cookie's life and decrease in value as the cookie ages.

Finally, other studies show reasons why cookie value remains constant over time. Recent studies by Neumann, Tucker, and Whitfield (2019), Neumann, Tucker, Subramanyam, and Marshall (2022), and Kraft, Miller, and Skiera (2023) suggest that consumer data obtained by online tracking technologies are inconsistent and sometimes even inaccurate. A study with UK managers (Weiss 2018) confirms this result: 82% believe that consumer data are somewhat unreliable. Consequently, inconsistent and inaccurate data represent a challenge for targeted advertising, as they may lead to wrongly personalized ads, a misunderstanding of consumer habits, and an erroneous assessment of ad effectiveness (Lucker, Hogan, and Bischoff 2017). The industry might anticipate inaccurate or inconsistent data in consumer profiles and ignore them in setting ad prices.

In summary, we know little about the value of cookies, specifically concerning cookie lifetime, cookie value, and the evolution of the value over time. Having such knowledge would make it easier to have a more thoughtful debate on the economic value of cookies and the economic consequences a restriction of cookie lifetimes might entail for the online advertising industry. Web Appendix W1 outlines the widely varying planned and already implemented initiatives to restrict cookie lifetimes currently under consideration across Europe.

## 5. Description of Empirical Study

*5.1 Aim of Empirical Study*

The goal of our empirical study is to determine the economic consequences of online tracking restrictions for publishers' revenues, specifically the following three:

1. Short cookie lifetime restriction regime (30 days, 60 days, 90 days, 120 days)

2. Medium cookie lifetime restriction regime (360 days)

3. Long cookie lifetime restriction regime (720 days)

These restrictions could be easily implemented by simply requiring the cookie owner to set the cookie's expiration date accordingly. The browser would automatically delete the cookie after this expiration date and, thus, provide a practical implementation of the right to be forgotten as outlined, for example, in Europe's GDPR.

*5.2 Description of Data*

Our data come from a large European ad exchange that is a programmatic marketplace for online display, mobile, and video advertising. The ad exchange reaches approximately 84% of its relevant market's total monthly internet users. This exchange granted us full access to a sample of its proprietary data for 867 days between March 3, 2014, and July 16, 2016, representing ~2.5 years of data on approximately 128 million ad impressions sold in a real-time auction to 54,127 cookies. Our data include desktop and mobile browsing traffic and are anonymized, so individuals remain unidentifiable.

Trading on the ad exchange occurs via real-time bidding (RTB; see Figure 3 and, e.g., Tunuguntla and Hoban 2020). Each observation represents an auction from a single ad impression a publisher sold on the ad exchange. We observe the third-party cookie of the ad exchange, the only cookie that the ad exchange sets. It proxies the advertiser's third-party cookie. The cookies of the ad exchange and the advertiser for a particular user overlap when the advertiser has won an auction for the particular user because serving an ad to the user allows the advertiser to place its third-party cookie in the user's browser. To our knowledge, this article is the first to exploit longitudinal auction data over ~2.5 years to gauge the value of a cookie and determine the economic consequences of cookie lifetime restrictions.

< INSERT FIGURE 3 ABOUT HERE >

*5.3 Description of Approach to Analyze Data*

Figure 4 describes our approach to quantifying the economic consequences of cookie lifetime restrictions for publishers.

< INSERT FIGURE 4 ABOUT HERE >

In Step 1, we draw a random sample of cookies, respectively, the user behind each cookie, from the RTB data set. The sample is a random sample of 54,127 users from a randomly selected day (i.e., Wednesday, April 29, 2015) at the center of the data set. We gathered all available information for these users for our entire observational period of ~2.5 years, resulting in 128,334,068 auctions that each sell one ad impression. The random sample represents 1% of unique cookies on our sampling day. We drew a sample that covers one day to get a representative sample of cookies because it (1) draws cookies from users who access the internet at different times during the day and (2) limits the likelihood that users who often delete their cookies were included more than once in our sample. Although sampling from one randomly chosen day may oversample heavy internet users because they are more likely active on that particular day, their activity on the specific day itself did not increase their probability of being included in the sample because we drew a random sample from all users of the particular day, not from all ad impressions.

In Step 2, we construct a longitudinal cookie panel by extracting all available cookie data per unique cookie identifier over the observation period. In Step 3, we adjust for potential censoring of cookie lifetime because the cookie might have lived longer than our observation period. Using a parametric survival model, we predict the remaining (unobserved) cookie lifetime (beyond our observation period) (for validation of this approach, see Web Appendix W3). In Step 4, we use regression analysis to uncover the incremental value per day (for validation of this approach, see Web Appendix W4). In Step 5, we use the obtained information to simulate the economic consequences of the various cookie lifetime restriction regimes. This approach is easily scalable and applicable to other real-time bidding data sets using tracking technologies.

*5.4 Description of Sample*

In the following, we describe the behavior of two selected cookies and then continue by computing summary statistics for the 54,127 cookies we observed in the sample. We describe separate analyses for cookie lifetime, cookie value per lifetime unit (e.g., days), and cookie lifetime value. Table 2 provides an overview of two selected cookies, and Table 3 gives an overview of our sample.

< INSERT TABLE 2 AND TABLE 3 ABOUT HERE >

*A detailed description of two cookies.* We find a positive relationship between cookie price and cookie lifetime for the cookie with the ID 177'239'342'526'XXX'XXX (see Figure 5). On our sampling day (April 29, 2015), the cookie was 149 days old. However, we observed the cookie in our longitudinal data from December 1, 2014, to July 15, 2016, which means that the cookie lived for another 443 days after our sampling day, yielding an observed lifetime of 592 days. During this time, the cookie was active on 514 days (86.824% of the observed days) and received 9,162 ad impressions. Thus, on average, 15.476 ad impressions per observed day.

Given that we observed the cookie last within seven days of the end of our observation period (July 16, 2016), we assume that the cookie lived longer than our observation period. Using a survival model, we predict that the cookie lived for another 481 days. Therefore, the uncensored lifetime of this cookie is 1,073 days. The observed mean price per ad impression was €1.181 CPM. The observed LVC (i.e., the sum of the prices paid by the advertisers) amounts to €10.823, giving us an observed average cookie value per day of €.018. The predicted censored LVC is €12.248, yielding an absolute percentage error (APE) of 13.166%, which corresponds to the in-sample absolute difference between the observed and the predicted lifetime value of the cookie divided by the observed lifetime value of the cookie (= abs(€10.823 − €12.248) / €10.823)). This cookie's predicted residual lifetime value is €17.687, resulting in an uncensored lifetime value of €29.935 (= €12.248 + €17.687) and an uncensored average value of the cookie per day of €.028 (= €29.935 / 1,073 days). Table 2 summarizes the descriptive statistics for this cookie.

In contrast, we find a negative relationship between cookie price and cookie lifetime for the cookie with the ID 466'830'604'730'XXX'XXX (see Figure 5). On our sampling day (April 29, 2015), the cookie had an age of 104 days. However, we observe the cookie in our longitudinal data from January 15, 2015, to September 11, 2015, which amounts to an observed cookie lifetime of 239 days. During this time, the cookie was active for 150 days (62.762 % of the observed days) and received 3,627 ad impressions (15.176 on average per observed day). Given that we observed the cookie last about ten months before the end of our observation period (July 16, 2016), we assume that we observed the entire lifetime of this cookie. The observed mean price per ad impression was €.554 CPM. The observed LVC (i.e., the sum of the prices paid by the advertisers) amounts to €2.011, giving us an observed average cookie value per day of .008 (= € 2.011 / 239 days). The predicted censored and, in this case, also uncensored lifetime value of this cookie is €2.366 (APE: 17.653%), which yields an uncensored average cookie value per day of €.010 (= €2.366 / 239 days). Table 2 summarizes the descriptive statistics for this cookie.

< INSERT FIGURE 5 ABOUT HERE >

*Summary statistics of the sample.* Our descriptive analysis of the data, as summarized in Table 3, shows a mean cookie age on our sampling day of 105 days (median: 16 days). We also find a mean cookie lifetime of 216 days (median: 68 days).

To account for potential censoring of observed cookie lifetime, we use a survival model to compute the predicted residual mean lifetime for those cookies that received ad impressions (and thus were active) within the first and last seven days of our observation period. Based on this cutoff criterion, 13.123% of all cookies are (potentially) censored: 5.572% are left-censored (i.e., we are only able to observe cookie death), 5.271% are right-censored (i.e., we are only able to observe cookie birth), and 2.280% are both right- and left-censored (i.e., we observe neither cookie birth nor cookie death).

We use the "flexsurv" package in R to determine the probability density for the death of a cookie at time t:

$$f(t|\mu, \alpha), t \geq 0,$$

where $\mu$ is the primary parameter of interest, which governs the mean or location of the underlying distribution, and $\alpha$ determines the variance or shape of the distribution. We do not consider any additional covariates in this analysis. Precisely, we fit a parametric Weibull and parametric Lognormal survival model to the data, consisting of the observed lifetime per cookie and an indicator variable of whether the observed lifetime per cookie is censored, as defined above.

Our main analysis uses a seven-day cut-off criterion to define whether a cookie is censored. The choice of the threshold determines the type-1 error, i.e., considering a cookie is dead although it is still alive (e.g. because it would appear again after eight days). So, choosing a low threshold, such as our seven days, yields an economic loss of cookie lifetime restrictions that is lower than if we would choose a more extended threshold. However, increasing the threshold to 28 days only slightly increases our loss estimates, providing confidence that the choice of the threshold does not impact our results too strongly (see Web Appendix W3 for details).

Furthermore, we only use cookies with an observed cookie lifetime of more than seven days to fit the survival models. Otherwise, we mix up very short lifetimes with very long lifetimes, and the short lifetimes will not impact our simulation results. This restriction reduces our sample size for the survival analysis to 32,189 cookies. We report the resulting model parameters and fit measures in Table 4. We select the Weibull model for further analysis because it fits the data better (Weibull LL: −179,085.300, Akaike information criterion [AIC]: 358,174.600, Bayesian information criterion [BIC]: 358,191.400 vs. Lognormal LL: −179,653.600, AIC: 359,311.300, BIC: 359,328.000). We validate this approach in Web Appendix W3.

We find the residual mean cookie lifetime $T$ per cookie $i$ for all 13.123% censored cookies by following Meeker and Escobar's (1998) approach:

$$T_i(x_i) = \frac{\int_{x_i}^{\infty} S(t_i) dt_i}{S(x_i)} = \frac{\int_{x_i}^{\infty} \exp(-(t_i/\mu)^{\alpha}) dt_i}{\exp((-(x_i/\mu)^{\alpha})},$$

where $x_i$ corresponds to the observed and potentially censored lifetime of cookie $i$, $S(t_i) = \exp(-(t_i/\mu)^{\alpha})$ corresponds to the survival function of the Weibull distribution that yields the probability that cookie $i$ will survive beyond a specific time $t_i$, $\mu$ is the location parameter and $\alpha$ the shape parameter of the Weibull distribution. The resulting mean cookie lifetime adjusted for censoring is 279 days (median: 68 days).

The cookies of our sample were active on average on 74 of the observed days (median: 8 days), yielding an average share of observed active days at the observed days of 60.500% (median:

68.400%). The cookies also differ regarding the number of ad impressions served. They reach an average number of ad impressions of 2,371 (median: 38) and an average number of ad impressions per observed day of 12.582 (median: 2.333).

We calculate that cookie values per day in our data set reach a mean value of €.006 (median €.002). We also compute the average uncensored cookie value per day at €.007 (median: €.002). Finally, the observed mean value per 1,000 ad impressions paid by the purchasing advertiser is €.696 CPM, and the median price is €.643 CPM. For the observed cookie lifetime value, we find a mean lifetime value of €1.428 (median: €.022). The mean predicted censored lifetime value is €1.622 (median: €.022). We calculate a mean APE (MAPE) of .082 (median: .000), corresponding to the in-sample absolute difference between the observed lifetime value of the cookie and its predicted lifetime value divided by its observed lifetime value. The mean predicted residual lifetime value amounts to €.856 (median: .000). The uncensored cookie lifetime value yields a mean value of €2.522 (median: €.022).

< INSERT TABLE 4 ABOUT HERE >

## 6. Simulation Study

### 6.1 Setup of Simulation Study

We use a numerical example to outline the procedure of our simulation study to determine the economic loss associated with a restriction of cookie lifetime. In this example, we look at three cookies, each receiving one ad impression per day.

Cookie A lives for 22 (active) days. The price of the ad impression for this cookie is $.09 on the first day. It increases by $.01 on each additional day. This price measures the value of the cookie for the publisher. It is $.09 on day 1, $.10 on day 2, $.11 on day 3, and so forth, until $.30 on day 22. The LVC is thus the sum of the cookie values per day across the 22 days ($.09 + $.10 + $.11 +… + $.30 = $4.29).

What would be the economic loss of a restriction of this cookie's lifetime to ten days (i.e., 45% of its original lifetime)? There is no economic loss for the first ten days. After ten days, the cookie is deleted, but a new cookie is born. We assume that the user does not change her browsing behavior because of the automatic deletion of the cookie due to the cookie lifetime restriction, in line with Drèze and Zufryden's (1998) finding that tracking cookies' presence or absence does not influence users' browsing behavior. Furthermore, we assume that the user consents again to being tracked.

So, the new cookie will have a value on its first day equal to the value of the old cookie on its first day—the same holds for the second day, the third day, and so forth. Thus, for the next ten days, the new cookie has a value per day that is 10 × $.01 = $.10 per day lower than the old cookie. After 20 days, including ten days of the new cookie, the new cookie is deleted and reborn with a value on its first day equal to the value of the old cookie on its first day. Thus, for the last two days of the cookie's life, days 21 and 22, the reborn cookie has a value per day 20 × $.01 = $.20 lower than the old cookie. Overall, the lifetime restriction of the cookie leads to an economic loss of 10 days times $.10 per day and 2 days times $.20 per day, or $1.40. As a result, LVC decreases by $1.40, from $4.29 to $2.89 (−33%).

This decrease diminishes publishers' revenue if the advertisers move these savings to other (more efficient) advertising media (e.g., publishers that allow targeting based on user logins). If, however, advertisers operated with fixed budgets and aimed to spend the entire budget assigned to online display advertising, they would start to spend the savings on other online display ads whose prices would increase. In such a case, the losses for publishers would be lower. Therefore, we consider our calculated values to be the upper bounds of an economic loss to publishers.

The numerical example outlines that we can easily derive the daily increase in the cookie value by estimating a regression with the value per day as a dependent variable and time (here: days) as an independent variable. The respective result for this regression is value per day = $.08 + $.01 × day.

Thus, the time parameter (i.e., the parameter for the day) represents the value increase per day of a cookie, and the constant (here: $.08) essentially represents the value of a cookie that is independent of time (here: days). It reflects the value of a user without a cookie.

The value of the cookie beyond the constant essentially reflects the incremental value that builds over time. It reflects the cookie's value because it enables the advertiser to track, profile, segment, and target users across multiple websites and determine each impression's effectiveness. The positive value corresponds to the notion that older cookies are more valuable to the advertising industry because they contain more information, leading to higher prices in real-time bidding advertising auctions (e.g., Beales and Eisenach 2014; Casale 2015; Neumann, Tucker, and Whitfield 2019). The value of the constant reflects that advertisers also derive value from ads that do not require tracking users over time.

Another cookie in the numerical example, Cookie B, lives for 10 days. The ad impressions for this cookie always sell at $.19 per day, such that the value per day is unaffected by time, and the LVC is $1.90 (= 10 × $.19). The regression for this cookie yields a constant of $.19 and a time parameter of zero. As the value of this cookie does not increase over time, restricting the cookie's lifetime will not affect the value per day and, thus, the lifetime value of the cookie. Such a case could occur when advertisers do not value a user's growing browsing history over time but are, for example, merely interested in advertising in attractive contexts.

Our third cookie, Cookie C, has a lifetime of 18 days. The price of its ad impression on the first day is $.29 and decreases by $.01 for each additional day. Thus, the cookie generates a value of $.29 on day 1, $.28 on day 2, $27 on day 3, and so forth, and, thus, $.12 on day 18. The LVC is the sum of the daily cookie values ($.29 + $.28 + $.27 + … + $.12) = $3.69.

In contrast to Cookie A, this cookie would not incur an economic loss but a gain due to a cookie lifetime restriction of ten days (which represents only 56% of its original lifetime). The reason could be that the advertiser learns that the user has less value (e.g., by not purchasing despite clicking on ads). The lifetime restriction of this cookie now leads to an economic loss equal to the sum of 8 days × −$.10 per day, thus −$.80. As a result, LVC is increased by $.80 from $3.69 to $4.49 (+22%).

If the study only involved those three cookies, then the overall results would be that the sum of LVC without the lifetime restriction is $9.88 (= $4.29 + $1.90 + $3.69) and $9.28 (= $2.89 + $1.90 + $4.49) with the lifetime restriction, so $.60 lower. The average LVC decreased from $3.29 to $3.09 (−$.20). Thus, the restriction in cookie lifetime to ten days causes a total economic loss of $.60 and a loss per cookie of $.20 (6%).

Table 5 summarizes these results, as does Table 7 for our simulation analysis of the 54,127 cookies in our sample (Panel 1). Web Appendix W5 illustrates how our simulation study accommodates that (1) the predicted lifetime is longer than the observed lifetime, (2) cookies are not active every day, and (3) eliminates differences in the daily number of impressions per cookie across time.

< INSERT TABLE 5 ABOUT HERE >

## 6.2 Results of Regression Analysis

We use the procedure described for the numerical example to determine the economic loss with a restriction of cookie lifetime ranging between 30 and 720 days. In its first step, this procedure requires running a regression for each cookie that determines the change in value per time unit (e.g., day). In its second step, we use the results of all regressions to derive the economic loss in the simulation analysis.

*Setup of regression analysis.* We propose the following regression to model the incremental effect of time on the value of the cookie:

$$VALUE_{i,t} = \beta_{i,0} + \beta_{i,1}\ DAYCOUNT_{i,t} + \beta_{i,2}\ ADINVENTORY_{i,t} + \varepsilon_{i,t}\ ,(1)$$

where $VALUE_{i,t}$ is the value of the cookie $i$ on day $t$ (here measured by the average price per ad impression on day $t$). $DAYCOUNT_{i,t}$ is a variable with a value of 1 on the first day that the cookie lives, 2 on the second day, 3 on the third day, and so forth. The parameter of this variable is our time parameter that reflects the incremental effect of time (here: day) on the value of the cookie (here: the average of the prices per ad impression per day). Ultimately, this parameter captures the value of behavioral targeting for advertisers because it is a proxy for the value of the information collected on a specific cookie over time. We consider the incremental effect as positive (negative) if the value of the time parameter is positive (negative) and significant (at a 1% level). If the value is insignificant, we can conclude that there is no incremental effect.

As advertisers do not only purchase ads on our ad exchange based on the capacity to conduct behavioral targeting, we add control variables for $ADINVENTORY_{i,t}$ characteristics (i.e., media type, which captures the share of video ads over regular display ads per day; fold position, which captures the share of ads displayed above the fold per day; and the share of retargeted ad impressions per day). $\varepsilon_{i,t}$ is the error term.

Given that we estimate one regression per user, our regression analysis accounts for time-invariant individual differences across users, that is, user-fixed effects (e.g., operating system, browser type, browser language). We regress separately for each cookie the value per day (measured by the average price for all ad impressions by the advertisers on that day) on the variable $DAYCOUNT_{i,t}$. Thus, we run an ordinary least-squares (OLS) regression for each cookie (54,127 cookies in our sample).

We validate this approach using the logarithmic values for our dependent variable and different sets of independent variables. Web Appendix W4 reports the estimates of our cookie-specific regressions. It shows, for example, that our preferred model 2, in which we regress the average price per ad impression per day on day count and additional ad inventory characteristics, yields the best MAPE, which corresponds to the in-sample absolute difference between the observed cookie's lifetime value and its predicted lifetime value divided by its observed lifetime value in our sample (mean: .175, median: .100). In addition, it shows that our preferred model 2 obtains a better fit than model 1 (which does not consider ad inventory characteristics), as measured by AIC (model 1 mean: 236.820 vs. model 2 mean: 230.750; model 1 median: 106.460 vs. model 2 median:101.780) and BIC (model 1 mean: 243.970 vs. model 2 mean: 240.380; model 1 median: 112.780 vs. model 2 median: 110.350). Concerning model fit, the log-linear model (model 4) and the quadratic model (model 5) have a slightly better R-squared than model 2 but have a worse (model 4) or comparable (model 5) Akaike information criterion (AIC) and Bayesian information criterion (BIC).

*Linear model without additional covariates.* We summarize the results of our regression analysis in Table 6. In model 1, we find a positive incremental effect for 6,719 cookies. They represent 12.413% of all cookies in our sample. They received 48.309% of all ad impressions, with an average number of ad impressions per cookie of 9,228. Their mean uncensored cookie lifetime is 628 days, and the average uncensored cookie's lifetime value is €9.965.

< INSERT TABLE 6 ABOUT HERE >

We find a significant negative incremental effect for 4,357 cookies, representing 8.050% of all cookies in our sample. They received 17.956% of all ad impressions, with an average number of ad impressions per cookie of 5,289. Their mean cookie lifetime is 610 days, and the average uncensored cookie lifetime value is €5.175.

For most cookies (43,051, i.e., 79.537%), the time parameter is insignificant, i.e., zero. Specifically, 10,987 cookies, or 20.299% of all cookies, have this "zero effect". Nevertheless, these cookies received 33.739% of all ad impressions, with an average number of ad impressions per cookie of 1,006. Their mean uncensored cookie lifetime is 191 days, and the average uncensored cookie lifetime value is €1.032.

*Linear model with additional covariates.* In model 2, our preferred model specification (see Equation 1), we find a positive incremental effect for 6,932 cookies. They represent 12.807% of all cookies in our sample. They received 53.153% of all ad impressions, with an average number of ad impressions per cookie of 9,841. Their mean uncensored cookie lifetime is about 633 days, and the average uncensored cookie's lifetime value is €10.530.

We find a significant negative incremental effect for 4,109 cookies, representing 7.591% of all cookies in our sample. They received 15.030% of all ad impressions, with an average number of ad impressions per cookie of 4,695. Their mean uncensored cookie lifetime is 608 days, and the average uncensored cookie lifetime value is €4.876.

The time parameter is insignificant for most cookies (43,086 or 79.602%). Specifically, 10,951 cookies, or 20.232% of all cookies, have a zero effect in the sample. Nevertheless, these cookies received 31.817% of all ad impressions, with an average number of ad impressions per cookie of 948. Their mean uncensored cookie lifetime is 190 days, and the average uncensored cookie lifetime value is €1.010.

We provide the regression parameter plots for our primary model (model 2 in Table 6) in Figure 6. We report the distribution of the constant (Min = .000, Median = .000, Mean = .409, Max = 2.220) as well as the distribution of the time parameter (Min = −.007, Median = .000, Mean = .000, Max = .016). In Web Appendix W6, Table W6.1, we additionally show the results of two regressions to describe which user-level characteristics explain the constants and time parameters. These descriptive regressions show, for example, that the constant is significantly larger for cookies from a national cookie (i.e., a user from the country of our ad exchange) using a mobile device, a Chrome operating system, and a Chrome browser. It also shows that the time parameter is significantly lower for cookies with a BlackBerry or Chrome operating system.

< INSERT FIGURE 6 ABOUT HERE >

*6.3 Results of Simulation Study*

We next evaluate the impact of different restrictions on the lifetime of cookies. Such a restriction only affects the value of a cookie if the cookie fulfills two conditions. First, the cookie's lifetime is longer than the restriction of the cookie lifetime (condition I). Second, the time parameter of the regression is either significantly positive (condition II) or significantly negative (condition III). As outlined in our numerical example, cookies with a non-significant increase in value over time will not impact our loss (gains) calculations.

Table 7 reports the shares of cookies that fulfill either conditions I and II or conditions I and III under different cookie lifetime restriction policies and the share of the lost value of all cookies. A 30-day restriction will affect 54.773% of our sample cookies. Of those, 22.410% also have a positive incremental value per day (which represents 12.275% of all cookies in our sample). These cookies have an average lifetime value of €10.966, and they lose, on average, €4.135 or 37.707% of their average lifetime value. Further, 13.769% have a negative incremental value per day (which represents 7.542% of all cookies in our sample). These cookies have an average lifetime value of €4.906, and their loss is, on average, −€2.082 or −42.434% of their average lifetime value. The respective economic loss for all cookies in our sample with an average lifetime value of a cookie of €2.522 is €.351 or 13.916%. In Table 7, we also report the results of a 60, 90, and 120-day cookie lifetime restriction, which leads to an average percentage loss of 13.638%, 13.083%, and 12.647%, respectively.

< INSERT TABLE 7 ABOUT HERE >

The 360-day restriction advocated by the European Union (Article 29 Data Protection Working Party 2010) would affect 28.320% of cookies in our sample. Of those cookies, 28.234% also have a positive incremental value per day, such that the policy affects 7.996% of all cookies. These cookies have an average lifetime value of €15.332, and their loss is, on average, €3.785 or 24.686%. Further, 16.413% of cookies have a negative incremental value per day, representing 4.648% of cookies in our sample. These cookies have an average lifetime value of €7.025, and their loss is, on average, −€1.883, respectively −26.800%. The respective economic loss for all cookies in our sample with an average lifetime value of a cookie of €2.522 is €.215 or 8.524%.

Finally, a 720-day restriction, as some industry players such as Facebook have implemented, would affect 14.693% of cookies in our sample. Of those cookies, 35.986% also have a positive incremental value per day. Thus, the policy affects about 5.288% of cookies. These cookies have an average lifetime value of €20.401, and their loss is, on average, €3.796 or 18.607%. Further, 17.805% of cookies have a negative incremental value per day, representing 2.616% of cookies in our sample. These cookies have an average lifetime value of €9.863, and their loss is, on average, −€2.454 or −24.877%. The economic loss for all cookies in our sample with an average lifetime value of €2.522 is €.137 or 5.432%.

*6.4 Robustness of Results of Simulation Study*

To examine the robustness of the results, we use the same approach as for the first sample to select a second sample. For this second sample, we randomly drew 44,400 users from another randomly selected day (i.e., Friday, June 12, 2015) at the center of the data set. We again gathered all available information for these users for our entire observational period. This selection leaves us with 105,198,803 auctions that each sell one ad impression. The random sample represents 1% of all cookies on our sampling day.

We obtain similar results for our second sample (see Table 7, Panel 2). For example, the resulting average percentage loss of a 30-day cookie lifetime restriction is 13.861%, 13.148% for a 60-day restriction, 12.719% for a 90-day restriction, 12.290% for a 120-day restriction, 8.324% for a 360-day restriction, and 5.394% for a 720-day restriction. Overall, the results of the second sample are statistically indistinguishable from the first sample, which supports the robustness of our results. The detailed results from the second sample are available in Web Appendix W7.

## 7. Summary of Results and Implications

Tracking technologies such as cookies track what users do online, making them a central concern of many privacy debates. However, little is known about cookies, particularly their lifetime, their value, and the evolution of their value over time. This lack of knowledge makes a profound discussion about cookie restrictions challenging, which is unsatisfying for regulators and the advertising industry. Methodologically, we develop an approach to quantifying the economic loss of lifetime restrictions to online tracking, which applies to other data sets that contain ad prices (e.g., data sets from other ad exchanges). Our approach can also apply to restrictions of other tracking technologies, such as digital fingerprinting and login data.

Our analysis of 54,127 cookies covering approximately 128 million ad impressions over ~2.5 years shows a mean uncensored cookie lifetime of 279 days (9.3 months) and a median of 68 days (2.3 months). On average, 45.227% of cookies are deleted within one month, 49.081% within two months, 52.440% within three months, 55.307% within four months, 71.680% within 12 months, and 85.307% within 24 months. This finding indicates that a restriction of cookie lifetime to 12 months, as suggested, for example, by the European Union for the upcoming ePrivacy Regulation, and is, for example, already implemented in Italy, would affect 28.320% of all cookies (as those cookies live longer than 12 months). A restriction to 24 months, as implemented in Spain and by advertising industry stakeholders (e.g., Facebook), would affect only 14.693% of the cookies.

Our simulation study reveals that the economic loss to publishers of restricting cookie lifetimes to one year, as the European Union proposes (Council of the European Union 2021), is 8.524% of the cookies' total lifetime value. In light of the €10.6 billion cookie-based display ad revenues in the European Union (Interactive Advertising Bureau 2017), such a policy would incur a yearly loss of approximately €904 million in display ad revenues, which is equivalent to a yearly economic loss per EU internet user of €2.082 (see Table 8 for an overview). A two-year restriction would cut 5.432% of the cookies' lifetime value. Such a restriction policy would affect €576 million of the EU cookie-based yearly display ad revenues and incur a yearly economic loss per EU internet user of €1.327.

< INSERT TABLE 8 ABOUT HERE >

Our results contrast Chiou and Tucker (2021), who find no significant impact of a shorter data retention policy on the accuracy of search engine results. However, many searches are unique and outline precisely what the user intends to find. So, historical data may be of less value in search than in display advertising.

So, what are the implications for the marketing strategies of publishers? First, publishers learn that cookie lifetime restrictions of 12 or 24 months will not affect most users because they delete their cookies within 12 (71.680%) and 24 (85.307%) months. Therefore, any proposed or adopted regulation that provides longer lifetimes for cookies is unlikely to have any substantive impact on many publishers and users. Second, publishers learn about the size of the economic loss of providing more privacy, which is about a €2 yearly loss per user in case of a one-year cookie lifetime restriction. So, publishers would be indifferent between (i) a shorter lifetime of cookies and a subscription of all users at a yearly price of €2 or (ii) an unlimited lifetime and a free subscription. Publishers could also invest more decisive in alternative targeting strategies, e.g., contextual targeting. In discussions with regulators, they can use the €2 yearly loss per user to discuss a compensation value for providing more privacy. Such a compensation value could also be in the users' interest because it would enable publishers to provide higher-quality content (Shiller, Waldfogel, and Ryan 2018). Alternatively, publishers could offer users a share of their advertising revenues if they consent to be tracked for a more extended (or infinite) period.

Additionally, publishers could offer users more choices when asking for their consent. Currently, if publishers ask for consent, as they must in the European Union, they usually offer two alternatives: providing or denying consent. Instead, they could ask users to consent for a specific period, e.g., 1, 3, 6, 12, or 24 months. Our results show that more extended consent periods are better for publishers, but the more granular choices might incentivize users to deny their consent less often.

Our results indicate to advertisers and publishers that about 80% of cookies do not change in value over time. So, increasing their privacy by not collecting data over time does not come with an economic loss. It might be a suitable compromise between the industry and the regulator not to track these users. Other forms than behavioral targeting, e.g., contextual targeting, might be more appropriate for those users.

The loss of the remaining 20% of users adds up to a yearly loss of about €2 per user. So, there is a price for privacy, and the question for the regulator is: Is this price justified, and who should pay for this economic loss? Justifying the price requires comparing the economic loss with the (non-economic) gain of more privacy. We cannot answer this question because our cost-benefit analysis only covers the costs. However, future research could study the benefits, i.e., whether users' yearly benefit, measured by their willingness to pay, is above those €2. However, it is difficult to determine the extent to which users value privacy, given that there is a significant disparity between users' stated and revealed preferences regarding privacy, a disparity referred to as the "privacy paradox".

If the regulator did not want publishers to suffer from such a loss, then either users or taxpayers would have to compensate for this loss. Our results also outline that cookie lifetime restrictions, or, more generally, limiting the period a tracker can track a user, could represent a compromise between a complete ban of tracking and tracking a user for an infinite period. Our study reveals that regulators should consider shorter cookie lifetime restrictions to better protect user privacy, as almost half of all cookies (45.227%) are deleted within the first 30 days. However, this would also imply larger losses for publishers.

Our results show that cookie lifetime restrictions represent an economically significant loss for publishers. It also shows that, for example, a cookie lifetime restriction to 12 months would incur a yearly loss of €904 million which represents about 1 percent of the online advertising market of €86 billion in Europe. However, not all of this money flows to publishers.

The users learn from our results that denying tracking entirely or beyond a certain period reduces publishers' profit. The question is whether users should care. Our empirical study only partly enables us to answer this question. We show that economic losses exist and are not minor. However, we did not show whether publishers can bear them or have to react. Users could suffer from publishers' reactions, such as creating paywalls, increasing their prices, or decreasing the quality of their content.

Finally, we note that tracking technologies such as cookies may also provide value for users. Cookies enable publishers to personalize their content, which benefits users. For example, websites use cookies to identify users, remember their preferences, set up personalized content, and help users complete tasks without re-entering information when they revisit a website. Usually, a first-party cookie is sufficient for accomplishing this personalization. Furthermore, cookies enable advertisers to better target the user with ads, rendering their ads more effective. This better targeting can also benefit the user, who receives more relevant ads to help her make better purchase decisions. In some circumstances, however, targeted ads can also be associated with lower-quality vendors and higher prices for identical products than untargeted ads (Mustri, Adjerid, and Acquisti 2023).

European Union (2017b), "Proposal for a Regulation of the European Parliament and the Council Concerning the Respect for Private Life and the Protection of Personal Data in Electronic Communications and Repealing Directive 2002/58/EC (Regulation on Privacy and Electronic Communications)," Council of the European Union (September 8), http://data.consilium.europa.eu/doc/document/ST-11995-2017-INIT/en/pdf.

Garante per la Protezione dei Dati Personali (2015), "Cookies Instructions Kit," Italian Data Protection Authority (June 2), https://help.iubenda.com/wp-content/uploads/2018/04/Cookie-Law-Official-Kit-en.pdf.

Goldfarb, Avi and Catherine E. Tucker (2011a), "Privacy Regulation and Online Advertising," *Management Science,* 57 (1), 57–71.

Goldfarb, Avi and Catherine E. Tucker (2011b), "Online Display Advertising: Targeting and Obtrusiveness," *Marketing Science*, 30 (3), 389–404.

Google (2022a), "Expanding Testing for the Privacy Sandbox for the Web," (July 27), https://blog.google/products/chrome/update-testing-privacy-sandbox-web/.

Google (2022b), "Get to know the new Topics API for Privacy Sandbox," (January 25), https://blog.google/products/chrome/get-know-new-topics-api-privacy-sandbox/.

Google (2022c), "How Google Retains Data That We Collect" (accessed January 29, 2022), https://policies.google.com/technologies/retention.

Gross, Ralph and Alessandro Acquisti (2005), "Information Revelation and Privacy in Online Social Networks," in *Proceedings of the 2005 ACM Workshop on Privacy in the Electronic Society*. New York: Association for Computing Machinery, 71–80.

Hoban, Paul R. and Randolph E. Bucklin (2015), "Effects of Internet Display Advertising in the Purchase Funnel: Model-Based Insights from a Randomized Field Experiment," *Journal of Marketing Research*, 52 (3), 375–93.

HubSpot (2023), "State of Marketing Report 2023,", https://offers.hubspot.com/thank-you/state-of-marketing.

Interactive Advertising Bureau (2017), "ADEX Benchmark 2016: European Online Advertising Spend," https://www.iabeurope.eu/research-thought-leadership/resources/iab-europe-report-adex-benchmark-2016-the-definitive-guide-to-europes-online-advertising-market.

Interactive Advertising Bureau (2023a), "IAB Internet Advertising Revenue Report," https://www.iab.com/wp-content/uploads/2023/04/IAB_PwC_Internet_Advertising_Revenue_Report_2022.pdf.

Interactive Advertising Bureau (2023b), "IAB Europe AdEx Benchmark 2022 Study Reveals Strong Digital Advertising Growth," https://iabeurope.eu/all-news/iab-europe-adex-benchmark-2022-study-reveals-strong-digital-advertising-growth.

Jerath, Kinshuk, and Klaus M. Miller (2023), "Perceived Consumer Privacy Violations Under Different Firm Practices in Online Advertising," Working Paper.

Jin, Yuxi, and Bernd Skiera (2022), "How do Privacy Laws Impact the Value for Actors in the Online Advertising Market? A Comparison of the EU, US, and China," *Journal of Creating Value*, 8 (2), 306-327.

Johnson, Garrett A. (2013), "The Impact of Privacy Policy on the Auction Market for Online Display Advertising," Working Paper.

Johnson, Garrett A. (2023), "Economic Research on Privacy Regulation: Lessons from the GDPR and Beyond," NBER Working Paper.

Johnson, Garrett A., Randall A. Lewis, and Elmar I. Nubbemeyer (2017), "Ghost Ads: Improving the Economics of Measuring Online Ad Effectiveness," *Journal of Marketing Research*, 54 (6), 867–84.
39

# Tables

TABLE 1
OBSERVATIONAL AND EXPERIMENTAL STUDIES ON COOKIES

| Study | Aim | Observations | | | Cookie Information | | | |
|---|---|---|---|---|---|---|---|---|
| | | Period | Cookies | Ad Impressions | Age | Lifetime | Value | Value over Time |
| Abraham, Meierhoefer, and Lipsman (2007) | Determine Impact of Cookie Deletion on Website- and Ad-Metrics | 1 month | 400'000 | n.a. | ≤ 1 month [a] | no | no | no |
| Goldfarb and Tucker (2011a) | Determine Impact of Privacy Regulation on Ad Effectiveness | 55 days | n.a. | n.a. | no | no | 65% | no |
| Johnson (2013) | Determine Impact of Privacy Regulation on Ad Industry | 1 week | n.a. | n.a. | no | no | 39%-46% | no |
| Beales and Eisenach (2014) | Determine Value of a Cookie for Advertisers | 2 weeks | n.a. | 3.0 Mio. | 1.5 months [b] | no | $.22-$.34 premium | ↑ $.02-$.15 per month |
| Aziz and Telang (2016) | Determine Value of a Cookie for Advertisers | 1 day | 115'417 | 1.3 Mio. | no | no | no | no |
| Budak, Goel, Rao, and Zervas (2016) | Determine Impact of Ad Blocking and Tracking Restrictions | 1 year | 13.6 Mio. | n.a. | no | no | $2 | no |
| Johnson, Lewis, and Nubbemeyer (2017) | Improve Ad Effectiveness Measurement | 2 weeks | 566'377 | n.a. | 3.5 months [c] | no | no | no |
| Johnson, Shriver, and Du (2020) | Determine Effect of Cookie Opt-Out for Ad Industry | n.a. | n.a. | 62.9 Mio. | no | no | 52% | no |
| Marotta, Abhishek, and Acquisti (2019) | Determine Effect of Cookies on Publisher Revenue | 1 week | n.a. | n.a. | no | no | 8% | no |
| Ravichandran and Korula (2019) | Determine Effect of Cookies on Publisher Revenue | 96 days | n.a. | n.a. | no | no | 52% | no |
| Rafieian and Yoganarasimhan (2021) | Determine Value of Mobile Targeting | 30 days | 728,340 | 27 Mio. | no | no | 12% | no |
| Laub, Miller, and Skiera (2023) | Determine Effect of Cookies on Publisher Revenue | 2 weeks | 1.4 Mio. | 42.4 Mio | no | no | 18% | no |
| Wang, Jiang, Yang (2023) | Determine Effect of GDPR on Display Advertising | 10 weeks | n.a. | n.a. | no | no | 6% | no |
| **This study** | **Determine Economic Consequences of Online Tracking Restrictions** | **2.5 years** | **98,527** | **234 Mio.** | **3.5 months [b] 16 days [c]** | **9.3 month [b] 2.3 months [c]** | **€2.52 [b] €.02 [c]** | **↑12.807% ↓7.591% n.s. 79.602% [d]** |

Notes: [a] 31% of users delete their cookies within one month. [b] Mean. [c] Median. [d] For 12.807% of cookies, we find a significant positive incremental effect of time on the value of a cookie. For 7.591% of cookies, we find a significant negative incremental effect of time on the value of a cookie. For 79.602% of cookies, we find a non-significant incremental effect of time on the value of a cookie.

TABLE 2
DESCRIPTION OF BEHAVIOR OF TWO COOKIES

| Row | Cookie ID | 177'239'342'526'XXX'XXX | 466'830'604'730'XXX'XXX |
|---|---|---|---|
| I | Date and Time of First Impression | 2014-12-01 16:51:17 | 2015-01-15 14:41:29 |
| II | Date and Time of Last Impression | 2016-07-15 18:33:20 | 2015-09-11 10:49:39 |
| III | Observed Age of Cookie on Sampling Day (in days)[a] | 149 | 104 |
| IV = II − I | Observed (potentially censored) Lifetime of Cookie (in days) | 592 | 239 |
| V = f(IV) | Predicted Residual Lifetime of Cookie (in days) | 481 | 0 |
| VI = IV + V | Uncensored Lifetime of Cookie (in days)[b] | 1,073 | 239 |
| VII = f(IV) | Observed Number of Active Days | 514 | 150 |
| VIII = VII / IV | Share of Observed Active Days per Observed Days | 86.824% | 62.762% |
| IX | Observed Number of Ad Impressions | 9,162 | 3,627 |
| X = IX / IV | Observed Number of Ad Impressions per Day | 15.476 | 15.176 |
| XI = XIV / IV | Observed Value of Cookie per Day (in €) | .018 | .008 |
| XII = XVIII / VI | Uncensored Value of Cookie per Day (in €) | .028 | .010 |
| XIII = XIV / IX | Observed Value of Cookie per Ad Impression (in €, CPM) | 1.181 | .554 |
| XIV = f(II, VII)) | Observed (potentially censored) Lifetime Value of Cookie (in €) | 10.823 | 2.011 |
| XV = f(IV, XI) | Predicted Censored Lifetime Value of Cookie for Observed Lifetime (in €) | 12.248 | 2.366 |
| XVI = f(XIV, XV) | Absolute Percentage Error (APE) for Observed Lifetime | 13.166% | 17.653% |
| XVII = f(V, XI) | Predicted Residual Lifetime Value of Cookie for Residual Lifetime (in €) | 17.687 | .000 |
| XVIII = XV + XVII | Uncensored Lifetime Value of Cookie (in €)[c] | 29.935 | 2.366 |

Notes: [a] Rounded to the next full day. [b] We use a Weibull model to determine the predicted residual lifetime for 13.123% of the cookies with potentially censored cookie lifetime in sample 1. [c] We determine the uncensored cookie lifetime value using the regression from Equation 1 (i.e., model 2 in Table 6).

TABLE 3
SUMMARY STATISTICS PER COOKIE (N = 54,127)

| Category | Variable | Quantiles | | | | | Mean | SD |
|---|---|---|---|---|---|---|---|---|
| | | Min. | 25% | 50% | 75% | Max. | | |
| Lifetime Unit of Cookie | Observed Age of Cookie on Sampling Day (in days)[a] | 1 | 1 | 16 | 168 | 423 | 105 | 147 |
| | Observed (potentially censored) Lifetime of Cookie (in days)[a] | 1 | 1 | 68 | 416 | 867 | 216 | 268 |
| | Uncensored Lifetime of Cookie (in days)[a,b] | 1 | 1 | 68 | 416 | 1,351 | 279 | 396 |
| | Observed Number of Active Days | 1 | 1 | 8 | 73 | 836 | 74 | 138 |
| | Share of Observed Active Days per Observed Days | .004 | .181 | .684 | 1.000 | 1.000 | .605 | .391 |
| | Observed Number of Ad Impressions[c] | 0 | 4 | 38 | 771 | 539,264 | 2,371 | 10,569 |
| | Observed Number of Ad Impressions per Day | .000 | .878 | 2.333 | 9.000 | 4,176 | 12.582 | 55.347 |
| Value of Cookie per Lifetime Unit | Observed Value of Cookie per Day (in €) | .000 | .000 | .002 | .006 | .990 | .006 | .015 |
| | Uncensored Value of Cookie per Day (in €) | .000 | .000 | .002 | .007 | .990 | .007 | .017 |
| | Observed Value of Cookie per Ad Impression (in €, CPM) | .000 | .340 | .643 | .924 | 113.890 | .696 | .829 |
| Lifetime Value of Cookie | Observed (potentially censored) Lifetime Value of Cookie (in €) | .000 | .003 | .022 | .498 | 331.048 | 1.428 | 5.234 |
| | Predicted Censored Lifetime Value of Cookie for Observed Lifetime (in €) | .000 | .003 | .022 | .556 | 398.183 | 1.622 | 6.007 |
| | Mean Absolute Percentage Error (MAPE[d]) for Observed Lifetime | .000 | .000 | .000 | .091 | 11.222 | .082 | .229 |
| | Predicted Residual Lifetime Value of Cookie for Residual Lifetime (in €) | .000 | .000 | .000 | .000 | 238.448 | .856 | 5.114 |
| | Uncensored Lifetime Value of Cookie (in €)[e] | .000 | .003 | .022 | .610 | 449.403 | 2.522 | 10.603 |

Note: [a] Rounded to the next full day. [b] We use a Weibull model to determine the expected residual lifetime for 13.123% of the cookies with potentially censored cookie lifetime. The uncensored average cookie lifetime of 279 days is 29.167% larger than the average observed cookie lifetime of 216 days in the data. [c] 202 cookies were active on our sampling day but received no ad impressions. [d] MAPE corresponds to the in-sample absolute difference between the cookie's observed lifetime value and the cookie's uncensored lifetime value divided by the cookie's observed lifetime value. [e] We determine the uncensored cookie lifetime value using the regression from Equation 1 (i.e., model 2 in Table 6).

TABLE 4
SURVIVAL MODEL PARAMETERS AND FIT MEASURES

| Model | Shape Parameter α [95%-CI] | SE | Scale Parameter μ [95%-CI] | SE | LL | AIC | BIC |
|---|---|---|---|---|---|---|---|
| | | | N = 54,127 | | | | |
| Weibull | .979 [.969;.990] | .005 | 463.321 [457.501;469.215] | 2.988 | −179,085.300 | 358,174.600 | 358,191.400 |
| Lognormal | 5.633 [5.617;5.480] | .008 | 1.389 [1.380;1.402] | .006 | −179,653.600 | 359,311.300 | 359,328.000 |

Notes: We only consider cookies with an observed cookie lifetime of more than seven days to fit the survival models resulting in a sample size of 32,189. CI: confidence interval; SE: standard error; LL: loglikelihood value; AIC: Akaike information criterion; BIC: Bayesian information criterion

TABLE 5
ECONOMIC LOSS OF A TEN-DAY LIFETIME RESTRICTION IN OUR NUMERICAL EXAMPLE

| Cookie Lifetime Restriction | Fulfillment of Condition I[a] | | | | Fulfillment of Conditions I and II[b] | | Fulfillment of Conditions I and III[c] | | Economic Loss per Cookie that Fulfills Conditions I and II | | | Economic Loss per Cookie that Fulfills Conditions I and III | | | Economic Loss per Cookie | | |
|---|---|---|---|---|---|---|---|---|---|---|---|---|---|---|---|---|---|
| | No. Cookies | % (N = 3) | % Cond.II[d] | % Cond.III[e] | No Cookies | % (N = 3) | No Cookies | % (N = 3) | Average LVC | Average Absolute | Average % Loss | Average LVC | Average Absolute | Average % Loss | Average LVC | Average Absolute | Average % Loss |
| Ten days | 2 | 67% | 50% | 50% | 1 | 33% | 1 | 33% | $4.29 | $1.40 | 33% | $3.69 | −$.80 | −22% | $3.29 | $.20 | 6% |

Notes: [a] Condition I refers to the number of cookies with a cookie lifetime larger than the cookie lifetime restriction. Condition II refers to the number of cookies that increase in value per day. Condition III refers to the number of cookies that decrease in value per day. We consider the value increase (decrease) per day to be positive (negative) if the sign of the time parameter in our regression model is positive (negative) and the value of the parameter is significant (at a 1% level). [b] Conditions I and II refer to the number of cookies that fulfill condition I and increase their values per day. [c] Conditions I and III refer to the number of cookies that fulfill conditions I and decrease their values per day. [d] Share of those cookies that also fulfill condition II (P(Cond.II | Cond. I) = P(Cond. I & II) / P(Cond. I). [e] Share of those cookies that also fulfill condition III (P(Cond.III | Cond. I) = P(Cond. I & III) / P(Cond. I).

Reading example: 2 cookies (i.e., 67% of all cookies) fulfill condition I (i.e., have a cookie lifetime that is larger than the imposed cookie restriction of ten days), and 50% of these cookies either fulfill condition II or condition III. 1 cookie (i.e., 33% of all cookies) fulfills conditions I and II (i.e., increase in value per day). One cookie (i.e., 33% of all cookies) fulfills conditions I and III (i.e., decreases in value per day). The average cookie that fulfills conditions I and II has an average cookie lifetime value (LVC) of $4.29 and loses under a ten-day lifetime restriction an average of $1.40 (i.e., 33% of the total average LVC). The average cookie that fulfills conditions I and III has an LVC of $3.69 and loses under a ten-day lifetime restriction on average −$.80 (i.e., −22% of the total average LVC). The average cookie in our sample has an LVC of $3.29 and loses, on average, a value of $.20 (i.e., 6%) under a ten-day cookie restriction policy.

TABLE 6
REGRESSION RESULTS OF IMPACT OF TIME ON THE AVERAGE PRICE PER AD IMPRESSION PER DAY

|  | Significant Positive Incremental Effect | | Significant Negative Incremental Effect | | Nonsignificant Incremental Effect | |
| --- | --- | --- | --- | --- | --- | --- |
|  | Model 1 | Model 2 | Model 1 | Model 2 | Model 1 | Model 2 |
| Dependent Variable (in €; CPM) | Average Price per Ad Impression per Day | | | | | |
| Constant [95%- CI] | .500 [.332; .674] | .622 [.415; .845] | 1.146 [.943; 1.335] | 1.137 [.887; 1.359] | .000 [.000; .000] | .000 [.000; .000] |
| Time Parameter [95%- CI] | .001 [.000; .002] | .001 [.000; .002] | −.001 [−.002; −.000] | −.001 [−.002; −.000] | .000 [.000; .000] | .000 [.000; .000] |
| Additional Covariates | N | Y | N | Y | N | Y |
| Number of Cookies (% of all cookies) | 6,719 (12.413%) | 6,932 (12.807%) | 4,357 (8.050%) | 4,109 (7.591%) | 43,051 (79.537%) | 43,086 (79.602%) |
| Total Number of Ad Impressions (% of all) | 61,996,813 (48.309%) | 68,213,838 (53.153%) | 23,043,923 (17.956%) | 19,288,148 (15.030%) | 43,293,332 (33.739%) | 40,832,082 (31.817%) |
| Ad Impressions per Cookie | 9,228 | 9,841 | 5,289 | 4,695 | 1,006 | 948 |
| Mean (Median) Uncensored Cookie Lifetime (in days) | 628 (498) | 633 (504) | 610 (473) | 608 (468) | 191 (5) | 190 (6) |
| Mean (Median) Uncensored Cookie Lifetime Value (in €) | 9.965 (2.998) | 10.530 (3.078) | 5.175 (1.073) | 4.876 (1.058) | 1.032 (.009) | 1.010 (.009) |
| Number of Cookies with Significant Zero Effect (% of all cookies) | — | — | — | — | 10,987 (20.299%) | 10,951 (20.232%) |

Notes: Unless otherwise noted, this table reports our sample's median estimates from 54,127 cookie-specific regressions. We consider the value increase (decrease) per day to be positive (negative) if the sign of the time parameter is positive (negative) and the value of the parameter is significant (at a 1% level). If the value is insignificant, then we conclude that there is no increase (decrease) in value over time. We apply a small Winsorization to accommodate outliers and replace the most extreme values with the 99% quantile of the respective parameter estimate. Model 1 only includes the time parameter (here: day count) as the independent variable. Model 2 includes the time parameter (here: day count) and additional covariates (i.e., ad inventory characteristics) as independent variables.

TABLE 7
ECONOMIC LOSS OF VARIOUS COOKIE LIFETIME RESTRICTIONS

| Cookie Lifetime Restriction | Fulfillment of Condition I[a] | | | | Fulfillment of Conditions I & II[b] | | Fulfillment of Conditions I & III[c] | | Economic Loss per Cookie that Fulfills Conditions I & II | | | Economic Loss per Cookie that Fulfills Conditions I & III | | | Economic Loss per Cookie | | |
|---|---|---|---|---|---|---|---|---|---|---|---|---|---|---|---|---|---|
| | No. Cookies | % of all Cookies Survived (Deleted) | % Cond.II[d] | % Cond.III[e] | No Cookies | % of all Cookies | No Cookies | % of all Cookies) | Average LVC | Average Absolute [95%-CI] | Average % Loss [95%-CI] | Average LVC | Average Absolute [95%-CI] | Average % Loss [95%-CI] | Average LVC | Average Absolute [95%-CI] | Average % Loss [95%-CI] |
| | | | | | | | | | | | | | | | | | |
| Panel 1: Simulation results based on Sample 1 (N = 54,127): Regression of average price per ad impression per day on day count + additional covariates (Model 2) | | | | | | | | | | | | | | | | | |
| 30 days | 29,647 | 54.773% (45.227%) | 22.410% | 13.769% | 6,644 | 12.275% | 4,082 | 7.542% | €10.966 | €4.135 [4.074; 4.196] | 37.707% [37.149; 38.265] | €4.906 | −€2.082 [−2.122; −2.041] | −42.434% [−43.260; −41.609] | €2.522 | €.351 [.340; .361] | 13.916% [13.480; 14.312] |
| 60 days | 27,561 | 50.919% (49.081%) | 22.662% | 14.608% | 6,246 | 11.540% | 4,026 | 7.438% | €11.604 | €4.234 [4.163; 4.305] | 36.491% [35.879; 37.102] | €4.969 | −€1.947 [−1.986; −1.907] | −39.172% [−39.972; −38.373] | €2.522 | €.344 [.333; .355] | 13.638% [13.202; 14.074] |
| 90 days | 25,743 | 47.560% (52.440%) | 23.164% | 15.305% | 5,963 | 11.017% | 3,940 | 7.279% | €12.078 | €4.209 [4.139; 4.270] | 34.851% [34.272; 35.430] | €5.061 | −€1.830 [−1.870; −1.791] | −36.167% [−36.945; −35.389] | €2.522 | €.330 [.320; .341] | 13.083% [12.687; 13.519] |
| 120 days | 24,191 | 44.693% (55.307%) | 23.600% | 15.692% | 5,709 | 10.547% | 3,796 | 7.013% | €12.527 | €4.200 [4.130; 4.270] | 33.531% [32.971; 34.092] | €5.227 | −€1.771 [−1.812 −1.731] | −33.890% [−34.661; −33.119] | €2.522 | €.319 [.309; .329] | 12.647% [12.251;13.044] |
| 360 days | 15,329 | 28.320% (71.680%) | 28.234% | 16.413% | 4,328 | 7.996% | 2,516 | 4.648% | €15.332 | €3.785 [3.714; 3.856] | 24.686% [24.226; 25.146] | €7.025 | −€1.883 [−1.949; −1.817] | −26.800% [−27.737; −25.864] | €2.522 | €.215 [.206; .224] | 8.524% [8.167; 8.881] |
| 720 days | 7,953 | 14.693% (85.307%) | 35.986% | 17.805% | 2,862 | 5.288% | 1,416 | 2.616% | €20.401 | €3.796 [3.723; 3.869] | 18.607% [18.248; 18.966] | €9.863 | −€2.454 [−2.569; −2.339] | −24.877% [−26.044; −23.711] | €2.522 | €.137 [.130; .143] | 5.432% [5.154; 5.669] |

# TABLE 8 (ctd.)
# ECONOMIC LOSS OF VARIOUS COOKIE LIFETIME RESTRICTIONS

| Cookie Lifetime Restriction | Fulfillment of Condition I[a] | | | | Fulfillment of Conditions I & II[b] | | Fulfillment of Conditions I & III[c] | | Economic Loss per Cookie that Fulfills Conditions I & II | | | Economic Loss per Cookie that Fulfills Conditions I & III | | | Economic Loss per Cookie | | |
|---|---|---|---|---|---|---|---|---|---|---|---|---|---|---|---|---|---|
| | No. Cookies | % of all Cookies Survived (Deleted) | % Cond.II[d] | % Cond.III[e] | No Cookies | % of all Cookies | No Cookies | % of all Cookies) | Average LVC | Average Absolute [95%-CI] | Average % Loss [95%-CI] | Average LVC | Average Absolute [95%-CI] | Average % Loss [95%-CI] | Average LVC | Average Absolute [95%-CI] | Average % Loss [95%-CI] |
| | | | | | | | | | | | | | | | | | |
| *Panel 2: Simulation results based on Sample 2 (N = 44,400): Regression of average price per ad impression per day on day count + additional covariates (Model 2)* | | | | | | | | | | | | | | | | | |
| 30 days | 25,624 | 57.712% (42.288%) | 23.127% | 15.848% | 5,926 | 13.347% | 4,061 | 9.146% | €11.068 | €4.291 [4.223; 4.359] | 38.766% [38.152; 39.380] | €4.802 | €-2.016 [-2.057; -1.975] | -41.989% [-42.839; -41.138] | €2.799 | €.388 [.375; .401] | 13.861% [13.396; 14.325] |
| 60 days | 23,830 | 53.671% (46.329%) | 23.760% | 16.513% | 5,662 | 12.752% | 3,935 | 8.863% | €11.542 | €4.236 [4.165; 4.307] | 36.703% [36.089; 37.317]] | €4.940 | €-1.942 [-1.983; -1.902] | -39.319% [-40.142; -38.496] | €2.799 | €.368 [.355; .381] | 13.148% [12.683; 13.612] |
| 90 days | 22,358 | 50.356% (49.644%) | 24.179% | 17.014% | 5,406 | 12.176% | 3,804 | 8.568% | €12.022 | €4.231 [4.158; 4.304] | 35.191% [34.586; 35.796] | €5.078 | €-1.862 [-1.902; -1.822] | -36.665% [-37.459; -35.871] | €2.799 | €.356 [.343; .368] | 12.719% [12.254; 13.148] |
| 120 days | 21,081 | 47.480% (52.520%) | 24.534% | 17.433% | 5,172 | 11.649% | 3,675 | 8.277% | €12.465 | €4.225 [4.151; 4.298] | 33.893% [33.302; 34.484] | €5.230 | €-1.789 [-1.830; -1.749] | -34.213% [-34.988; -33.437] | €2.799 | .€344 [.332; .356] | 12.290% [11.861; 12.719] |
| 360 days | 13,313 | 29.984% (70.016%) | 28.288% | 17.832% | 3,766 | 8.482% | 2,374 | 5.347% | €15.859 | €3.957 [3.878; 4.036] | 24.951% [24.453; 25.448] | €7.198 | €-1.926 [-1.994; -1.859] | -26.764% [-27.705; -25.822] | €2.799 | €.233 [.222; 243] | 8.324% [7.931; 8.681] |
| 720 days | 6,981 | 15.723% (84.277%) | 35.625% | 19.267% | 2,487 | 5.601% | 1,345 | 3.029% | €21.321 | €3.992 [3.911; 4.073] | 18.722% [18.343; 19.102] | €9.936 | €-2.408 [-2.518; -2.299] | -24.239% [-25.341; -23.137] | €2.799 | €.151. [.143; .158] | 5.394% [5.108; 5.644] |

Notes: [a] Condition I refers to the number of cookies with a cookie lifetime larger than the cookie lifetime restriction. Condition II refers to the number of cookies that increase their values per day. Condition III refers to the number of cookies that decrease their values per day. [b] Conditions I and II refer to the number of cookies that fulfill condition I and increase their values per day. [c] Conditions I and III refer to the number of cookies that fulfill condition I and decrease their values per day. [d] Share of those cookies that also fulfill condition II (P(Cond.II | Cond. I) = P(Cond. I & II) / P(Cond. I). [e] Share of those cookies that also fulfill condition III (P(Cond.III | Cond. I) = P(Cond. I & III) / P(Cond. I).

Reading example: 29,647 cookies (i.e., 54.773% of all cookies survived the imposed cookie lifetime restriction of 30 days whereas 45.227% were deleted) ) fulfill condition I (i.e., have a cookie lifetime larger than the imposed cookie lifetime restriction of 30 days), and 22.410% of these cookies fulfill condition II and 13.769% condition III. 6,644 cookies (i.e., 12.275% of all cookies) fulfill conditions I and II (i.e., increase in value per day). 4,082 cookies (i.e., 7.542% of all cookies) fulfill conditions I and III (i.e., decrease in value per day). The average cookie that fulfills conditions I and II has an average cookie lifetime value (LVC) of €10.966 and loses under a 30-day lifetime restriction on average €4.135 (or 37.707% of the total average LVC). The average cookie that fulfills conditions I and III has an average LVC of €4.906 and loses under a 30-day lifetime restriction on average −€.2.082 (i.e., −42.434% of the total average LVC). The average cookie in our sample has an LVC of €2.522 and loses, on average, a value of €.351 (or 13.916%) under a 30-day cookie lifetime restriction policy.

TABLE 9

ECONOMIC LOSS PER EU INTERNET USER DUE TO COOKIE LIFETIME RESTRICTIONS

| Cookie Lifetime Restriction | | 30 days | 60 days | 90 days | 120 days | 360 days | 720 days |
|---|---|---|---|---|---|---|---|
| Average % Loss for all Cookies [95% Confidence Interval] | Sample 1 | 13.916% [13.480; 14.312] | 13.638% [13.202; 14.074] | 13.083% [12.687; 13.519] | 12.647% [12.251; 13.044] | 8.524% [8.167; 8.881] | 5.432% [5.154; 5.669] |
| | Sample 2 | 13.861% [13.396; 14.325] | 13.148% [12.683; 14.325] | 12.719% [12.254; 13.612] | 12.290% [11.861; 12.719] | 8.324% [7.931; 8.681] | 5.394% [5.108; 5.644] |
| | Average | 13.889% [13.438; 14.319] | 13.393% [12.943; 14.200] | 12.901% [12.471; 13.566] | 12.469% [12.056; 12.882] | 8.424% [8.049; 8.781] | 5.413% [5.389; 5.657] |
| Yearly EU Cookie-Based Display Ad Revenue | | €10.600 billion[a] | | | | | |
| Affected Yearly EU Cookie-based Display Ad Revenue [95%- Confidence Interval] | Sample 1 | €1.475 billion [1.429; 1.517] | €1.446 billion [1.399; 1.492] | €1.387 billion [1.345; 1.433] | €1.341 billion [1.299; 1.383] | €904 million [866; 941] | €576 million [546; 601] |
| | Sample 2 | €1.469 billion [1.420; 1.518] | €1.394 billion [1.344; 1.518] | €1.348 billion [1.299; 1.443] | €1.303 billion [1.257; 1.348] | €882 million [841; 920] | €572 million [541; 598] |
| | Average | €1.472 billion [1.424; 1.518] | €1.420 billion [1.372; 1.505] | €1.368 billion [1.322; 1.438] | €1.322 billion [1.278; 1.365] | €893 million [853; 931] | €574 million [544; 600] |
| EU Internet Users | | 434 million[b] | | | | | |
| Yearly EU Cookie-Based Display Ad Revenue per User | | €24.424 | | | | | |
| Yearly Economic Loss per EU Internet User [95% Confidence Interval] | Sample 1 | €3.399 [3.292; 3.496] | €3.331 [3.224; 3.437] | €3.195 [3.099; 3.302] | €3.089 [2.992; 3.186] | €2.082 [1.995; 2.169] | €1.327 [1.259; 1.385] |
| | Sample 2 | €3.385 [3.272; 3.499] | €3.211 [3.098; 3.499] | €3.106 [2.993; 3.325] | €3.002 [2.897; 3.106] | €2.033 [1.937; 2.120] | €1.317 [1.248; 1.378] |
| | Average | €3.392 [3.282; 3.497] | €3.271 [3.161; 3.468] | €3.151 [3.046; 3.313] | €3.045 [2.945; 3.146] | €2.057 [1.966; 2.145] | €1.322 [1.253; 1.382] |

Notes: [a] This value refers to the tracking-based ad revenue during the time of our study (Interactive Advertising Bureau 2017). Very likely, today's tracking-based ad revenue is larger. [b] This value refers to the number of EU internet users during the time of our study (Statista 2019).

# Figures

## FIGURE 1
### OVERVIEW OF EXEMPLARY PLANNED AND IMPLEMENTED INITIATIVES TO RESTRICT ONLINE TRACKING

| Initiator<br><br>Status | Policy Makers | Industry |
|---|---|---|
| **Implemented** | <ul><li>EU ePrivacy Directive (ePD) 2002</li><li>EU General Data Protection Regulation (GDPR) 2018 (Opt-in consent)</li><li>California Consumer Privacy Act (CCPA) 2018 (Opt-out consent)</li><li>Digital Service Tax 2020</li></ul> | <ul><li>2018: Apple SKAdNetwork (SKAN)</li><li>2018: netID</li><li>2018: Tracking-free Subscriptions</li><li>2019: Apple Intelligent Tracking Prevention (ITP)</li><li>2021: Apple App Tracking Transparency (ATT)</li><li>2022: Mozilla Total Cookie Protection (TCP)</li></ul> |
| **Planned** | <ul><li>EU ePrivacy Regulation (ePR) (Restrictions of Tracking Technologies' Lifetime)</li></ul> | <ul><li>Google Privacy Sandbox (e.g., FloC, FLEDGE, Aggregate & Conversion Measurement APIs, Trust Token API, Topics API)</li></ul> |

## FIGURE 2
### PLANNED AND IMPLEMENTED INITIATIVES TO RESTRICT THE LIFETIME OF TRACKING TECHNOLOGIES (COOKIES)

| Initiator<br><br>Status | Policy Makers | Industry[b] |
|---|---|---|
| **Implemented** | All cookies<ul><li>France: 6 months[a] (Commission Nationale de l'Informatique et des Libertés 2020)<br>Italy: 12 months (Garante per la Protezione dei Dati Personali 2015)</li><li>Spain: 24 months (Agencia Española de Protección de Datos 2020)</li><li>UK and GER: No specification, but shorter lifetime advocated (Voisin et al. 2021)</li></ul> | First-party cookies<ul><li>13 months (Tradelab 2019)</li><li>18 months: Google (2022c)</li><li>24 months: Facebook (Cook 2017)</li></ul>Third-party cookies<ul><li>Full block: Apple (Wilander 2020), Brave (2022), Mozilla (Englehardt and Marshall 2019), Microsoft (Barker and Murgia 2020)</li></ul> |
| **Planned** | All cookies<ul><li>EU: 6−12 months (European Union 2017a; 2017b)</li><li>EU: 12 months (Article 29 Data Protection Working Party 2010)</li><li>EU: 12 months (Council of the European Union 2021)</li></ul> | Third-party cookies<ul><li>Full block: Google (Google 2022a)</li></ul> |

Notes: [a] The lifespan of analytic cookies benefiting from the CNIL consent exemption must not exceed 13 months.
[b] The industry initiatives have a worldwide reach and thus apply to users from all countries.

FIGURE 3
ILLUSTRATION OF THE AUCTION PROCESS IN REAL-TIME BIDDING

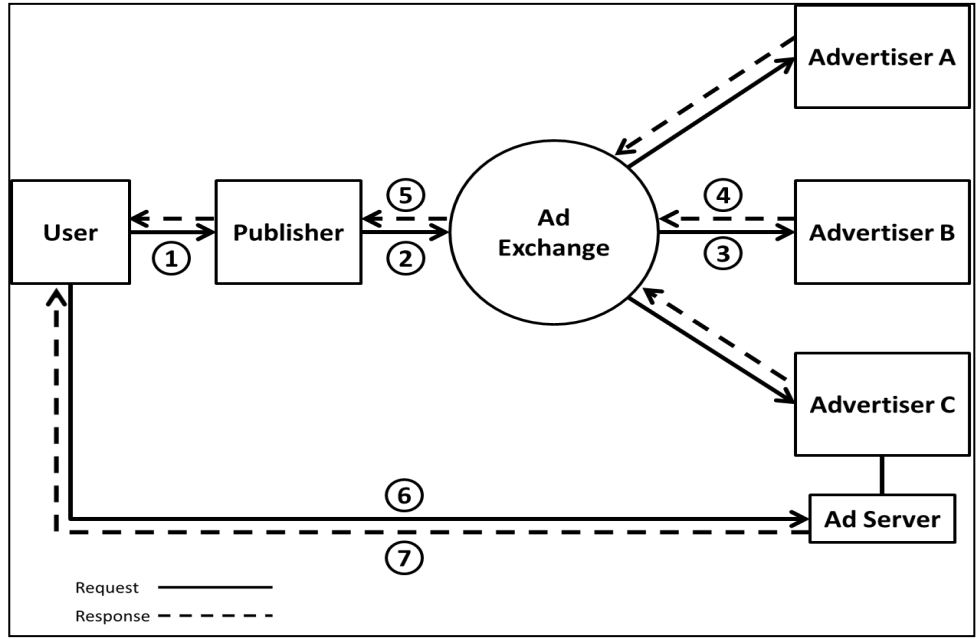

Notes: As shown in the figure, whenever a user visits a publisher's website with ad slots (1), the publisher sends an ad call to an ad exchange (2). This ad call is a request to run a real-time auction on the ad exchange and contains information about, for example, the properties of the ad slot (e.g., ad size) and a user ID, for example, a cookie ID, which we explain in more detail in Section 2.2. The ad exchange then sends a bid request to all advertisers on the ad exchange (3). Each interested advertiser submits a bid for displaying its ad to the user, including the ad server's address with the ad (4). The ad exchange determines the auction's price and winner and forwards this information to the publisher (5). The publisher then asks the user's browser to load the ad from the ad server (6) and displays the ad to the user on the publisher's website (7).

FIGURE 4
ILLUSTRATION OF APPROACH TO DERIVE ECONOMIC CONSEQUENCES
OF COOKIE LIFETIME RESTRICTIONS FROM REAL-TIME BIDDING DATA

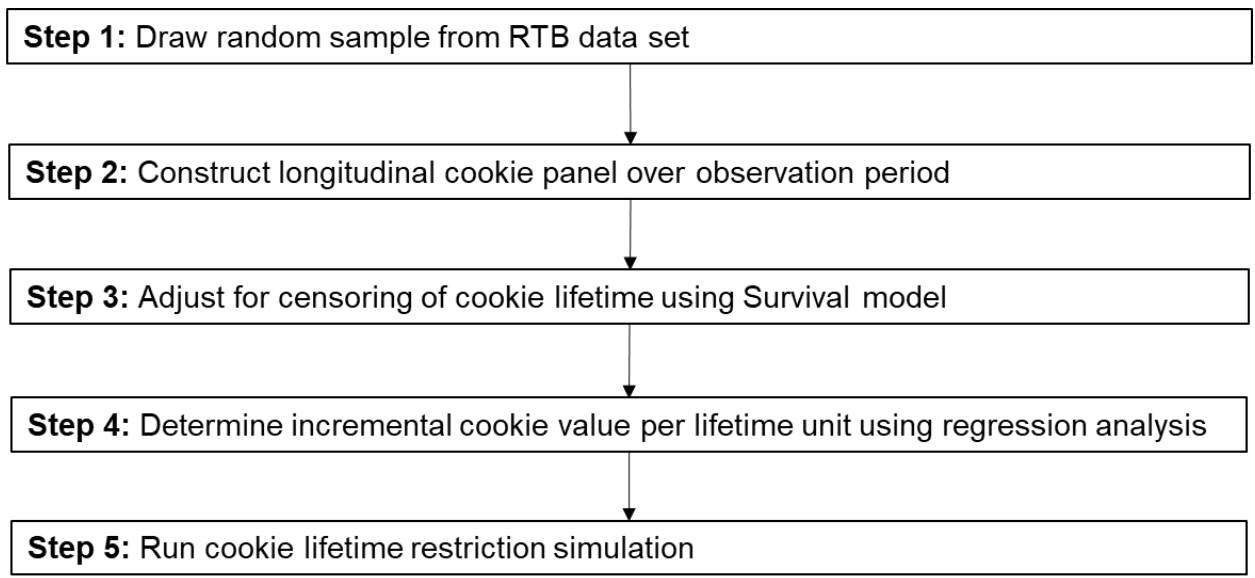

# FIGURE 5
## DEVELOPMENT OF AVERAGE PRICE PER AD IMPRESSION PER DAY OVER COOKIE LIFETIME FOR TWO EXEMPLARY COOKIES

Cookie ID 177'239'342'526'XXX'XXX 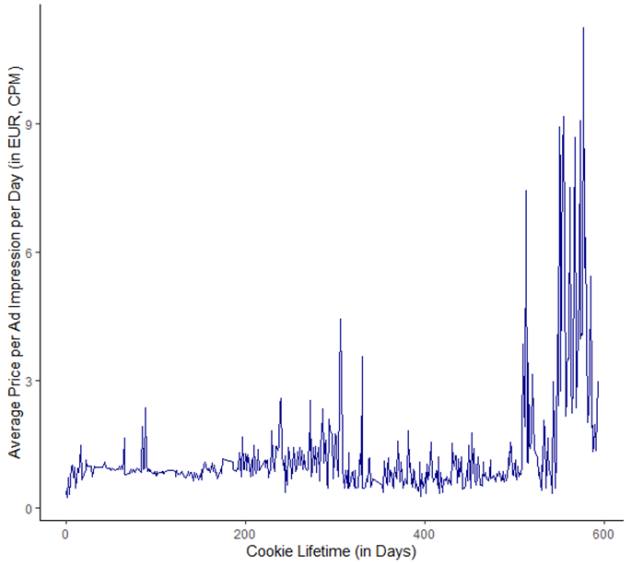  Cookie ID 466'830'604'730'XXX'XXX 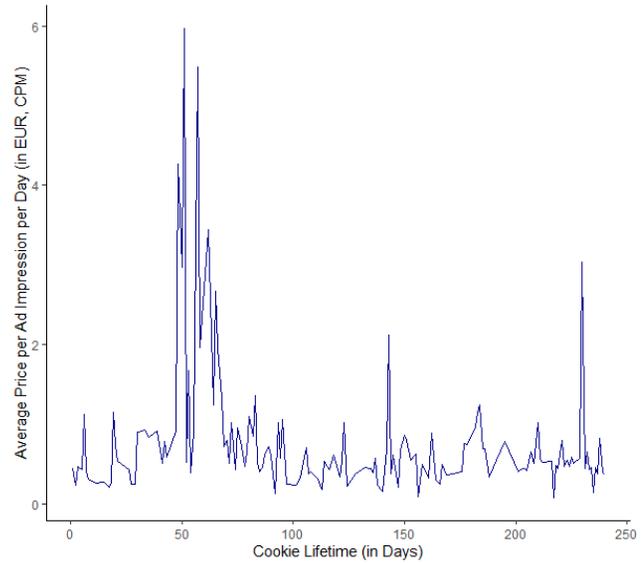

# FIGURE 6
## REGRESSION PARAMETER PLOTS

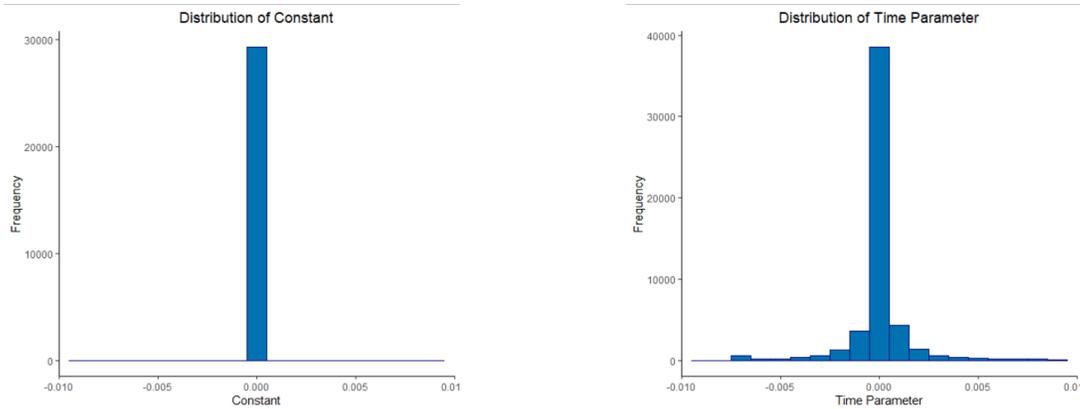

Notes: Dependent variable: average price per ad impression (in €, CPM). We apply a small Winsorization to accommodate outliers and replace the most extreme values with the 99% quantile of the respective parameter estimate. Distribution of constant: Min = .000, Median = .000, Mean = .409, Max = 2.220. Distribution of time parameter: Min = −.007, Median = .000, Mean = .000, Max = .016. Plots are restricted to [−.010; .010] for readability.

# Web Appendix W1
# Detailed Description of Initiatives of Policy Makers and the Online Advertising Industry to Restrict the Lifetime of Tracking Technologies

Online tracking in Europe is mainly governed by the ePrivacy Directive and the General Data Protection Regulation (GDPR). To date, these European laws have not restricted the lifetime of cookies or other tracking technologies. Still, some EU member states, such as Italy, France, and Spain, have already enacted such restrictions. The European Union is considering such restrictions on top of the current privacy laws, for example, within the upcoming ePrivacy Regulation (Council of the European Union 2021). This void in European regulation has led to various proposals on cookie lifetime and data retention periods from national data protection agencies and the online advertising industry, with unclear economic consequences for the online advertising industry.

As early as 2010, the Article 29 Data Protection Working Party (2010), now superseded by the European Data Protection Board (https://edpb.europa.eu/edpb_en), published an opinion on online behavioral advertising with some clarifications on how to interpret the ePrivacy Directive. The Working Party (p. 6) acknowledges that "cookies have different life spans. This lifespan might or might not be extended in the future upon further visits to the same site (this is a design decision by the programmer). 'Persistent cookies' either have a precise expiry date far in the future or until they are manually deleted."

Further, the Working Party (p. 16) outlines that "users' acceptance of a cookie could be understood to be valid not only for the sending of the cookie but also for subsequent collection of data arising from such a cookie. In other words, the consent obtained to place the cookie and use the information to send targeted advertising would cover subsequent 'readings' of the cookie every time the user visits a website partner of the ad network provider which initially placed the cookie. However, taking into account that i) this practice would mean that individuals accept to be monitored

'once for ever,' and, ii) individuals might simply 'forget' that, for example, a year ago, they agreed to be monitored; the Working Party considers that some safeguards should be implemented."

Specifically, the Working Party (p. 16) requires the online advertising industry "to limit the scope of the consent in terms of time. Consent to be monitored should not be 'for ever', but it should be valid for a limited period of time, for example, one year. After this period, ad network providers would need to obtain new consent. This setting could be achieved if cookies had a limited lifespan after they have been placed in the user's terminal equipment (and the expiry date should not be prolonged)." Article 29 Data Protection Working Party (2010; p. 24) adds, "Ad network providers should implement retention policies which ensure that information collected each time that a cookie is read is automatically deleted after a justified period of time (necessary for the purposes of the processing)." Article 29 Data Protection Working Party (2011; p. 7) specifies the following: "Furthermore, the collection and processing of data for behavioral advertising purposes must be kept to a minimum." Further, the Working Party (2011; pp. 7–8) criticizes that industry self-regulation efforts so far do "not contain any provisions on the amount of data collected and the retention period(s) for the specific purposes.… Such an initiative should at least address the period in which consent can be considered valid, and after which data shall then be deleted."

In 2013, Article 29 Data Protection Working Party (p. 3) further outlined the conditions for how to obtain valid consent from users: "Necessary information would be the purpose(s) of the cookies and, if relevant, an indication of possible cookies from third parties or third party access to data collected by the cookies on the website. Information such as the retention period (i.e., the cookie expiry date), typical values, details of third-party cookies and other technical information should also be included to fully inform users."

In 2015, Article 29 Working Party (p. 2) commissioned a cookie sweep of 478 websites in the e-commerce, media and public sectors across eight member states in the EU and "highlighted areas for improvement including a few cookies with duration periods of up nearly 8,000 years. This large value is in contrast to an average duration of 1 to 2 years. The sweep, however, also showed that 70% of the 16,555 cookies recorded were third-party cookies. Only 25 third-party domains set more than half of the third-party cookies." Article 29 Working Party (2015, p. 19) continues, "Some first[-] and third[-]party cookies appear to have an extremely long duration. Cookies with an expiry set to 31/12/9999 23:59 (the maximum possible value) could be regarded as not having a reasoned retention schedule defined. Excluding cookies with a long duration (greater than 100 years), the average duration was between 1 to 2 years. This could be a useful starting point for a discussion regarding an acceptable maximum duration, although the purpose of the cookie will also need to be taken into account."

Again, neither the ePrivacy Directive nor the GDPR contains an explicit time limit for the validity of consent, the appropriate maximum cookie lifetime and the retention period for the associated data. To fill this void in the existing legal framework, Article 29 Data Protection Working Party (2010, p. 16) recommends an overall cookie lifetime of not more than one year. The European Parliament has circulated draft versions of the new ePrivacy Regulation (ePR) with cookie lifetime requirements from six months to no longer than 12 months (European Union 2017a; 2017b; Council of the European Union 2021). The French data protection agency has already implemented a maximum cookie lifetime of six months in France (Commission Nationale de l'Informatique et des Libertés 2020)[1]. In contrast, the Italian data protection agency requires deleting cookies in Italy after 12 months (Garante per la Protezione dei Dati Personali 2015; p. 17), and the Spanish Agencia Española de Protección de Datos (2020; p. 29) requires deleting cookies after 24 months. In Germany and the UK, data protection authorities acknowledge cookie lifespan restrictions as necessary and advocate relatively shorter lifespans but do not specify how long the useful life of a cookie should be (Voisin et al. 2021).

---

[1] With the execption that the lifespan of analytic cookies benefitting from the CNIL consent exemption must not exceed 13 months.

In addition to the planned and already implemented initiatives of the various data protection authorities and regulatory bodies across Europe, various online advertising industry stakeholders have put forward their guidelines to restrict the lifetime of cookie tracking. For example, Tradelab, a sizable programmatic media buyer in Europe, has voluntarily committed to a maximum cookie lifetime of 13 non-rolling months and deletes the associated data after 12 months (Tradelab 2019). Google anonymizes advertising data by removing part of the IP address after nine months and their first-party cookie information after 18 months (Google 2022). Facebook keeps its first-party cookies for 24 months (Cook 2017). At the same time, prominent online advertising industry stakeholders such as Apple, Microsoft, Mozilla, and Google have either implemented or plan to completely ban third-party cookies (possibly also for the benefit of their first-party cookies) (Barker and Murgia 2020).

Furthermore, various cookie and online consent solutions claim to comply with the EU ePrivacy Directive and the GDPR, such as Cookiebot (2022), OneTrust (2022), Curac-Dahl and Juszczyński (2021), and Koch (2022). They allow a maximum cookie lifetime of 12 months (see also Sanchez-Rola et al. 2019).

# Web Appendix W2
# Exemplary Observed Maximum Cookie Lifetimes from Selected Domains

TABLE W2.1

EXEMPLARY OBSERVED MAXIMUM COOKIE LIFETIMES

FROM SELECTED DOMAINS

| Cookie Domain | Cookie Name (given by owner of cookie) | Expiration Date | Maximum Cookie Lifetime (in days) |
|---|---|---|---|
| amazon.com | _utma | March 13, 2025 | 730 |
|  | aws-at-main | February 9, 2043 | 7,272 |
|  | aws-priv | March 14, 2026 | 1,096 |
| google.com | anid | April 7, 2024 | 390 |
|  | apisid | March 13, 2025 | 730 |
|  | consent | October 19, 2041 | 6,794 |
| bing.com | muid | April 7, 2024 | 390 |
|  | srchid | April 7, 2024 | 390 |
|  | srchuid | April 7, 2024 | 390 |
| facebook.com | c_user | March 13, 2024 | 365 |
|  | datr | March 13, 2025 | 730 |
|  | dbln | March 13, 2025 | 730 |
| twitter.com | _ga | March 13, 2025 | 730 |
|  | ads_prefs | March 12, 2028 | 1,825 |
|  | dnt | March 12, 2028 | 1,825 |

Notes: Mean maximum cookie lifetime = 1,626 days (min = 365 days, max = 7,272 days). Cookies are generated from a single visit to the respective domains after deleting all old cookies and the entire browsing history on March 14, 2023, using Google Chrome (chrome://settings/siteData).

# Web Appendix W3
# Survival Model Validation

We use a third sample to validate our approach to determine the residual cookie lifetime for potentially censored cookies. To construct the sample, we focus on a cohort of new cookies to avoid any potential left-censoring and to be able to observe the actual cookie birthday in our data. We consider a cookie to be a member of this newborn cohort of cookies if the cookie is active for the first time in our focal sampling week but was not active at any time in the ten weeks before the beginning of our observation period. Thus, it is very likely that all cookies we observe for the first time in calendar week 20 (i.e., in the 11th week of our data set) are newborn cookies. This selection leaves us with 30,952 cookies in our subsample of newborn cookies. Only 18 cookies (.058%) are potentially right-censored (i.e., active within the last seven days of our observation period), but no cookies are left-censored. Thus, we have a maximum of 797 observable days for the newborn cookie cohort.

The third sample differs from the first and second samples as it includes only the newborn cookies in week 20 of our data. In contrast, samples 1 and 2 include cookies born on the respective sampling days and those born before our sampling days. It leads, for example, to a much larger observed cookie lifetime in samples 1 and 2 than in sample 3. This difference in composition of the different samples used in our study is, however, no reason for concern regarding the validation of our survival models used in our study. We use the newborn cohort of the third sample only for validation but not to obtain our main results.

Our descriptive analysis of the subsample of newborn cookies, as summarized in Table W3.1, shows a mean cookie age on our sampling day of 0 days, as expected. We find a mean cookie lifetime of 35 days (1.2 months) and a median of 1 day. The observed cookies differ concerning the number of ad impressions served. They reach an average number of ad impressions of about 142 and a median of four ad impressions. We calculate that cookie values per day in our data set reach a mean value of

€.006 and a median of €.001. Finally, the observed mean price per 1,000 ad impressions paid by the purchasing advertiser is €.807 CPM, and the median price is €.709 CPM. Concerning the observed cookie lifetime value, we find a mean lifetime value of €.095 and a median of €.004.

We split the newborn subsample into a training and test data set to validate our approach to determine residual cookie lifetime. We use the first 500 days (63%) as our estimation sample and estimate a parametric Weibull, Lognormal, and Generalized Gamma model. We report the model fit measures in Table W3.2 Panel 1. The Generalized Gamma model fits best, but the differences between all models are relatively small.

We use the remaining 297 days (37%) of our data as our holdout sample and report the respective validation measures in Panel 2 of Table W3.2. We follow Meeker and Escobar's (1998) suggestion to obtain the mean and median residual lifetime per cookie (RLT) as outlined in the main part of the paper. The Weibull model best predicts the observed mean cookie lifetime in our newborn sample of 34.100 [95% confidence interval (95% CI): 32.998; 35.203] with an estimated uncensored mean cookie lifetime of 35.380 [95% CI: 34.194; 36.566]. The Weibull model also shows the best validation measures with an R-squared between the observed and predicted values of cookie lifetime of .974, a mean absolute error (MAE) of 1.958, a root mean squared error (RMSE) of 18.221 and a mean absolute percentage error (MAPE) of .003.

Taken together, the parametric Weibull model provides the best fit. Figure W3.1 illustrates this conclusion by plotting the observed cookie lifetime and the Weibull model fit. Therefore, we subsequently use the parametric Weibull model to determine the uncensored mean cookie lifetime in samples 1 and 2 in the main part of the manuscript.

# TABLE W3.1

## SAMPLE 3: SUMMARY STATISTICS PER COOKIE – NEWBORN COHORT IN CALENDAR WEEK 20 OF 2014

(N = 30,952)

| Category | Variable | Quantiles | | | | | Mean | SD |
|---|---|---|---|---|---|---|---|---|
| | | Min. | 25% | 50% | 75% | Max. | | |
| Cookie Lifetime Units | Observed Cookie Age on Sampling Day (in days)[a] | 0 | 0 | 0 | 0 | 0 | 0 | 0 |
| | Observed Cookie Lifetime (in days)[ab] | 1 | 1 | 1 | 9 | 791 | 35 | 99 |
| | Observed Number of Ad Impressions | 1 | 1 | 4 | 18 | 87,313 | 142 | 1,497 |
| Cookie Value per Lifetime Unit | Observed Cookie Value per Day (in €) | .000 | .000 | .001 | .005 | 5.000 | .006 | .039 |
| | Observed Cookie Value per Ad Impression (in €, CPM) | .000 | 411 | .709 | 1.096 | 27.997 | .807 | .853 |
| Cookie Lifetime Value | Observed Cookie Lifetime Value (in €) | .000 | .001 | .004 | .018 | 56.059 | .095 | .908 |

Note: [a] Rounded to the next full day. [b] Using the data of sample 3, we generate a subsample of new cookies born in calendar week 20 in 2014. Our observation window starts in calendar week 10 in 2014, so we can use weeks 1–9 to check that the cookie was not observed before. This long period makes it very likely that all cookies observed for the first time in calendar week 20 are newborn cookies, so no cookies are left-censored. We have a maximum of 797 observable days for the newborn cookie cohort. Of these newborn cookies, only 18 cookies (.058% of all cookies in the subsample) are potentially right-censored (i.e., lived for more than 797 days).

TABLE W3.2

SAMPLE 3: SURVIVAL MODEL VALIDATION – NEWBORN COHORT IN CALENDAR WEEK 20 OF 2014 (N = 30,952)

| Model | Shape Parameter [95% CI] | SE | Scale Parameter [95%- CI] | SE | LL | AIC | BIC |
|---|---|---|---|---|---|---|---|
| *Panel 1: Model Fit Measures* | | | | | | | |
| Weibull | .823 [.808; .837] | .007 | 114.302 [111.136; 117.558] | 1.638 | −44,898.810 | 89,801.620 | 89,815.650 |
| Lognormal | 4.120 [4.093; 4.148] | .014 | 1.282 [1.262; 1.303] | .011 | −44,195.650 | 88,395.290 | 88,409.320 |
| Generalized Gamma | 3.664 [3.606; 3.722] | .030 | 1.190 [1.164; 1.216] | .013 | −44,039.840 | 88,085.670 | 88,106.710 |

| Model | Observed Mean Cookie Lifetime [95% CI] | Uncensored Mean Cookie Lifetime (Mean RLT)[a] [95%-CI] | R2 | MAE | RMSE | MAPE |
|---|---|---|---|---|---|---|
| *Panel 2: Validation Measures* | | | | | | |
| Weibull | | 35.380 [34.194; 36.566] | .974 | 1.958 | 18.221 | .003 |
| Lognormal | | 40.197 [38.643; 41.752] | .925 | 6.097 | 52.314 | .010 |
| Generalized Gamma | | 106.883 [99.472; 114.294] | .659 | 72.783 | 592.174 | .117 |
| Model | 34.100 [32.998; 35.203] | Uncensored Mean Cookie Lifetime (Median RLT)[b] [95%-CI] | R2 | MAE | RMSE | MAPE |
| Weibull | | 108.679 [106.424; 110.933] | .896 | 74.578 | 135.616 | 6.867 |
| Lognormal | | 70.230 [68.655; 71.805] | .939 | 36.130 | 63.039 | 3.891 |
| Generalized Gamma | | 53.027 [50.371; 55.683] | .827 | 18.926 | 155.187 | .030 |

Notes: To avoid cookies with very short lifetimes impacting our results too strongly, we only consider cookies with an observed cookie lifetime of seven or more days to predict residual cookie lifetime. Under the lognormal model, the shape parameter equals the mean, and the scale parameter equals the standard deviation. The generalized gamma model has an additional parameter kappa, which is k = −.742 [−.827; −.657], SE = .043. [a] We determine the uncensored mean cookie lifetime using the predicted mean residual lifetime (RLT). [b] We determine the uncensored mean cookie lifetime using the predicted median residual lifetime (RLT). SE: standard error; LL: log-likelihood; AIC: Akaike information criterion; BIC: Bayesian information criterion; RLT: residual lifetime; MAE: mean absolute error; RMSE: root mean squared error; MAPE: mean absolute percentage error.

FIGURE W3.1

SAMPLE 3: OBSERVED COOKIE LIFETIME AND WEIBULL MODEL FIT FOR NEWBORN COHORT IN CALENDAR WEEK 20 OF 2014 (N = 30,952)

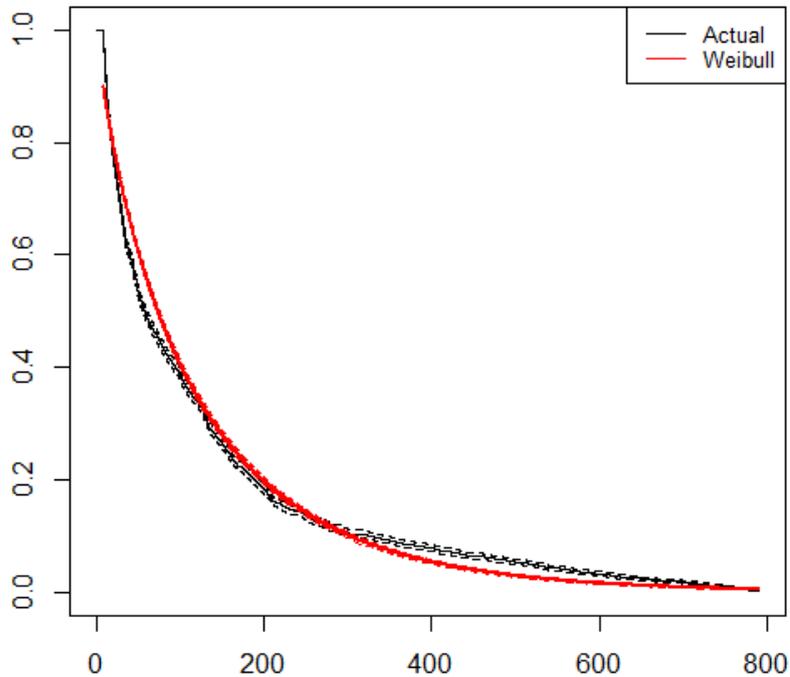

Another cause for concern is the 7-day threshold to determine whether we observe the birth or death of a cookie. In doing so, we are prone to two types of errors: First, we extend the observed lifetime of a cookie, although the cookie was actually deleted. Second, we do not extend the observed lifetime of a cookie, although the cookie lived longer than the observed lifetime.

If we choose a rather short threshold (i.e., seven days as in our main analysis), we are more likely to not extend the observed lifetime of a cookie, although the cookie lived longer. Alternatively, suppose we choose a rather long threshold (i.e., 28 days, as in our following robustness analysis). In that case, we are more likely to overestimate the lifetime of a cookie, although the cookie was deleted.

Our main specification in the paper uses a 7-day threshold to account for potential censoring of the observed cookie lifetime. Specifically, we use a survival model to compute the predicted residual mean lifetime for those cookies that received ad impressions (and thus were active) within the first and last seven days of our observation period. Based on this cut-off criterion, 13.123% of all cookies are (potentially) censored: 5.572% are left-censored (i.e., we are only able to observe cookie death), 5.271% are right-censored (i.e., we are only able to observe cookie birth), and 2.280% are both right- and left-censored (i.e., we observe neither cookie birth nor cookie death).

For robustness, we extend the threshold for user inactivity and compute the predicted residual mean lifetime for those cookies that received ad impressions (and thus were active) within the first and last 28 days of our observation period. Based on this cut-off criterion, 17.656% of all cookies are (potentially) censored. 7.015% are left-censored, 6.889% are right-censored, and 3.752% are right- and left-censored.

Using the 7-day (28-day) threshold, we obtain an average uncensored lifetime of a cookie of 279 (328) days and an uncensored lifetime value of a cookie of €2.522 (€2.958). That is, the uncensored lifetime of a cookie under a 28-day threshold is 17.563% larger than under the 7-day threshold, and the uncensored lifetime value of a cookie is 17.288% larger under the 28-day threshold than under the 7-day threshold (see Table W3.3).

Although we predict a larger lifetime and value of the cookies under the 28-day threshold, we only find a slightly larger economic loss of the various cookie lifetime restrictions (see Table W3.4). For example, under the 7-day (28-day) restriction, we find a 13.916% average %-loss (versus 15.888%) for the 30-day restriction. We also find a 5.432% average %-loss for the 720-day restriction (versus 7.403%). The differences in the average %-loss between the 7-day and the 28-day threshold are statistically significant, as the 95%-confidence intervals do not overlap.

In summary, our choice of a rather short threshold of seven days to determine censored observations in our data does not lead us to overestimate the economic loss of various cookie lifetime restrictions. Instead, our results provide a lower bound for economic loss. Increasing the threshold to 28 days only slightly increases our loss estimates, providing additional confidence in our results' robustness.

TABLE W3.3

ROBUSTNESS TO USER INACTIVITY THRESHOLD (N = 54,127)

| Threshold for User Inactivity | Variable | Quantiles | | | | | Mean | SD |
|---|---|---|---|---|---|---|---|---|
| | | Min. | 25% | 50% | 75% | Max. | | |
| - | Observed (potentially censored) Lifetime of Cookie (in days) [a] | 1 | 1 | 68 | 416 | 867 | 216 | 268 |
| | Observed (potentially censored) Lifetime Value of Cookie (in €) | .000 | .003 | .022 | .498 | 331.048 | 1.428 | 5.234 |
| 7 days | Uncensored Lifetime of Cookie (in days) [ab] | 1 | 1 | 68 | 416 | 1,351 | 279 | 396 |
| | Uncensored Lifetime Value of Cookie (in €) [c] | .000 | .003 | .022 | .610 | 449.403 | 2.522 | 10.603 |
| 28 days | Uncensored Lifetime of Cookie (in days) [ad] | 1 | 1 | 68 | 420 | 1,516 | 328 | 481 |
| | Uncensored Lifetime Value of Cookie (in €) [e] | .000 | .003 | .022 | .678 | 555.438 | 2.958 | 12.765 |

Note: [a] Rounded to the next full day. [b] We use a Weibull model to determine the expected residual lifetime for 13.123% of the cookies with potentially censored cookie lifetime and determine censored cookies based on a threshold of 7 days of inactivity from the beginning and the end of our observation period. The adjusted average predicted cookie lifetime of 279 days is 29.167% larger than the average observed cookie lifetime of 216 days in the data (i.e., sample 1). [c] We determine the uncensored cookie lifetime value using the regression from Equation 1 (i.e., model 2 in Table 7). [d] We use a Weibull model to determine the expected residual lifetime for 17.656% of the cookies with potentially censored cookie lifetime. We determine the censored cookies based on a threshold of 28 days of inactivity from the beginning and end of our observation period. The adjusted average predicted cookie lifetime of 328 days is 51.851% larger than the average observed cookie lifetime of 216 days in the data (i.e., sample 1) and 17.563% larger than the adjusted average predicted cookie lifetime of 279 days.

# TABLE W3.4

# ECONOMIC LOSS OF VARIOUS COOKIE LIFETIME RESTRICTIONS

| Cookie Lifetime Restriction | Fulfillment of Condition I[a] | | | | Fulfillment of Conditions I & II[b] | | Fulfillment of Conditions I & III[c] | | Economic Loss per Cookie that Fulfills Conditions I & II | | | Economic Loss per Cookie that Fulfills Conditions I & III | | | Economic Loss per Cookie | | |
|---|---|---|---|---|---|---|---|---|---|---|---|---|---|---|---|---|---|
| | No. Cookies | % of all Cookies | % Cond.II[d] | % Cond.III[e] | No Cookies | % of all Cookies | No Cookies | % of all Cookies) | Average LVC | Average Absolute [95%-CI] | Average % Loss [95%-CI] | Average LVC | Average Absolute [95%-CI] | Average % Loss [95%-CI] | Average LVC | Average Absolute [95%-CI] | Average % Loss [95%-CI] |
| *Panel 1: Simulation results based on Sample 1 (N = 54,127): Regression of average price per ad impression per day on day count + additional covariates (Model 2), 7 day threshold for user inactivity* | | | | | | | | | | | | | | | | | |
| 30 days | 29,647 | 54.773% | 22.410% | 13.769% | 6,644 | 12.275% | 4,082 | 7.542% | €10.966 | €4.135 [4.074; 4.196] | 37.707% [37.149; 38.265] | €4.906 | −€2.082 [−2.122; −2.041] | −42.434% [−43.260; −41.609] | €2.522 | €.351 [.340; .361] | 13.916% [13.480; 14.312] |
| 360 days | 15,329 | 28.320% | 28.234% | 16.413% | 4,328 | 7.996% | 2,516 | 4.648% | €15.332 | €3.785 [3.714; 3.856] | 24.686% [24.226; 25.146] | €7.025 | −€1.883 [−1.949; −1.817] | −26.800% [−27.737; −25.864] | €2.522 | €.215 [.206; .224] | 8.524% [8.167; 8.881] |
| 720 days | 7,953 | 14.693% | 35.986% | 17.805% | 2,862 | 5.288% | 1,416 | 2.616% | €20.401 | €3.796 [3.723; 3.869] | 18.607% [18.248; 18.966] | €9.863 | −€2.454 [−2.569; −2.339] | −24.877% [−26.044; −23.711] | €2.522 | €.137 [.130; .143] | 5.432% [5.154; 5.669] |
| *Panel 2: Simulation results based on Sample 1 (N = 54,127): Regression of average price per ad impression per day on day count + additional covariates (Model 2), 28 day threshold for user inactivity* | | | | | | | | | | | | | | | | | |
| 30 days | 29,647 | 54.773% | 22.410% | 13.769% | 6,644 | 12.275% | 4,082 | 7.542% | €13.413 | €5.362 [5.286; 5.438] | 39.976% [39.407; 40.546] | €5.236 | −€2.495 [−2.543; −2.447] | −47.651% [−48.571; −46.731] | €2.958 | €.470 [.457; .483] | 15.888% [15.449; 16.328] |
| 360 days | 15,329 | 28.320% | 28.234% | 16.413% | 4,328 | 7.996% | 2,516 | 4.648% | €19.089 | €5.607 [5.509; 5.705] | 29.374% [28.860; 29.887] | €7.561 | −€2.796 [−2.888; −2.703] | −36.974% [−38.195; −35.753] | €2.958 | €.318 [.306; .331]. | 10.750% [10.344; 11.189] |
| 720 days | 9,777 | 18.063% | 33.477% | 16.948% | 3,273 | 6.047% | 1,657 | 3.061% | €23.428 | €5.435 [5.350; 5.21] | 23.201% [22.837; 23.564] | €9.566 | −€3.578 [−3.710; −3.446] | −37.401% [−38.779; −36.024] | €2.958 | €.219 [.210;.228] | 7.403% [7.099; 7.708] |

Notes: [a] Condition I refers to the number of cookies with a cookie lifetime larger than the cookie lifetime restriction. Condition II refers to the number of cookies that increase their values per day. Condition III refers to the number of cookies that decrease their values per day. [b] Conditions I and II refer to the number of cookies that fulfill condition I and increase their values per day. [c] Conditions I and III refer to the number of cookies that fulfill condition I and decrease their values per day. [d] Share of those cookies that also fulfill condition II (P(Cond.II | Cond. I) = P(Cond. I & II) / P(Cond. I). [e] Share of those cookies that also fulfill condition III (P(Cond.III | Cond. I) = P(Cond. I & III) / P(Cond. I).

Reading example: 29,647 cookies (i.e., 54.773% of all cookies) fulfill condition I (i.e., have a cookie lifetime larger than the imposed cookie lifetime restriction of 30 days), and 22.410% of these cookies fulfill condition II and 13.769% condition III. 6,644 cookies (i.e., 12.275% of all cookies) fulfill conditions I and II (i.e., increase in value per day). 4,082 cookies (i.e., 7.542% of all cookies) fulfill conditions I and III (i.e., decrease in value per day). The average cookie that fulfills conditions I and II has an average cookie lifetime value (LVC) of €10.966 and loses under a 30-day lifetime restriction on average €4.135 (or 37.707% of the total average LVC). The average cookie that fulfills conditions I and III has an average LVC of €4.906 and loses under a 30-day lifetime restriction on average −$.2.082 (i.e., −42.434% of the total average LVC). The average cookie in our first sample has an LVC of $2.522 and loses, on average, a value of €.351 (or 13.916%) under a 30-day cookie lifetime restriction policy.

# Web Appendix W4
# Validation of Regressions

We run four other regression models to validate our regression results to determine the incremental cookie value per lifetime unit (Step 4 in Figure 3 in the main part of the manuscript).

In models 1 and 2, we use the average price per ad impression per day as the dependent variable. As an independent variable, we use the day count, which captures the incremental value of time. In model 1, we only use the day count variable as a covariate, while in model 2, we consider ad inventory characteristics as additional covariates. These ad inventory characteristics control for media type, which captures the share of video ads over regular display ads per day; fold position, which captures the share of ads displayed above the fold per day; and the share of retargeted ad impressions per day.

In models 3 and 4, we use ln(average price per ad impression per day) as the dependent variable. As an independent variable, we use day count without additional covariates (model 3) and day count and ad inventory characteristics as additional covariates (model 4).

Finally, in model 5, we use the average price per ad impression per day as the dependent variable. As an independent variable, we use day count, day count$^2$ and ad inventory characteristics as additional covariates.

Our main criterion for model selection is prediction quality (see Panel 1 in Table W4.1). Therefore, we calculate the prediction quality measures R-squared, mean average error (MAE), and the root mean squared error (RMSE) using the first 80% of consecutive observations of each cookie to train the model and the last 20% observations to test the model. Finally, we calculate the mean absolute percentage error (MAPE), corresponding to the in-sample absolute difference between the cookie's observed lifetime value and the cookie's predicted lifetime value divided by the cookie's observed lifetime value. Our preferred specification, the linear model with ad inventory characteristics as additional covariates (model 2), performs best on average, for example, with regard to MAPE. We, therefore, choose model 2 as our primary model in the main manuscript, which we use to determine the incremental cookie value per lifetime unit.

As model fit measures, we determine the R-squared, the Akaike information criterion (AIC), and the Bayesian information criterion (BIC) using the entire available data set (see Panel II in Table W4.1). With regard to model fit, the log-linear model (model 4) and the quadratic model (model 5) show a slight improvement in R-squared over model 2, but have a worse (model 4) or comparable fit (model 5) when comparing the Akaike information criterion (AIC), and the Bayesian information criterion (BIC).

We repeat the abovementioned analysis regarding our prediction quality and model fit measures for our second sample and obtain similar results. The detailed results are available from the authors upon request.

TABLE W4.1

ORDINARY LEAST SQUARES PREDICTION QUALITY AND MODEL FIT MEASURES

| Model Number | Dependent Variable | Independent Variable(s) | Measure | Quantiles | | | | | Mean | SD |
|---|---|---|---|---|---|---|---|---|---|---|
| | | | | Min | 25% | 50% | 75% | Max | | |
| | | | *Panel I: Prediction Quality Measures*[a] | | | | | | | |
| 1 | Average price per ad impression per day | Day count | $R^2$ | .000 | .029 | .117 | .367 | 1.000 | .261 | .318 |
| | | | MAE | .000 | .247 | .390 | .624 | 36.410 | .526 | .629 |
| | | | RMSE | .000 | .311 | .501 | .816 | 44.533 | .683 | .817 |
| | | | MAPE | .000 | .046 | .108 | .223 | 13.468 | .194 | .348 |
| 2 | Average price per ad impression per day | Day count ad inventory | $R^2$ | .000 | .025 | .116 | .425 | 1.000 | .274 | .330 |
| | | | MAE | .000 | .246 | .395 | .633 | 36.410 | .535 | .670 |
| | | | RMSE | .000 | .312 | .510 | .833 | 44.533 | .700 | .857 |
| | | | MAPE | .000 | .043 | .100 | .204 | 11.222 | .175 | .304 |
| 3 | ln (Averageprice per ad impression per day) | Day count | $R^2$ | .000 | .032 | .128 | .388 | 1.000 | .269 | .318 |
| | | | MAE | .001 | .357 | .485 | .671 | 16.937 | .592 | .502 |
| | | | RMSE | .001 | .446 | .600 | .809 | 17.018 | .706 | .541 |
| | | | MAPE | .000 | .061 | .138 | .263 | 10.814 | .202 | .246 |
| 4 | ln (Average price per ad impression per day) | Day count ad inventory | $R^2$ | .000 | .031 | .127 | .428 | 1.000 | .279 | .327 |
| | | | MAE | .001 | .355 | .484 | .670 | 11.991 | .594 | .509 |
| | | | RMSE | .001 | .444 | .598 | .810 | 16.177 | .711 | .560 |
| | | | MAPE | .000 | .052 | .120 | .240 | 41.495 | .185 | .338 |
| 5 | Average price per ad impression per day | Day count day count$^2$ ad inventory | $R^2$ | .000 | .028 | .117 | .399 | 1.000 | .269 | .325 |
| | | | MAE | .000 | .290 | .492 | .867 | 280.137 | .890 | 3.493 |
| | | | RMSE | .000 | .371 | .635 | 1.115 | 384.279 | 1.113 | 4.358 |
| | | | MAPE | .000 | .043 | .102 | .212 | 12.305 | .181 | .315 |

| Model Number | Dependent Variable | Independent Variable(s) | Measure | Quantiles | | | | | Mean | SD |
|---|---|---|---|---|---|---|---|---|---|---|
| | | | | Min | 25% | 50% | 75% | Max | | |
| | | | | Panel II: Model Fit Measures[b] | | | | | | |
| 1 | Average price per ad impression per day | Day count | $R^2$ | .000 | .006 | .028 | .088 | .953 | .073 | .113 |
| | | | AIC | −1,329.290 | 29.730 | 106.460 | 343.810 | 4,953.150 | 236.820 | 366.717 |
| | | | BIC | −1,321.630 | 33.180 | 112.780 | 353.910 | 4,967.200 | 243.970 | 369.104 |
| 2 | Average price per ad impression per day | Day count ad inventory | $R^2$ | .000 | .033 | .088 | .205 | .998 | .151 | .168 |
| | | | AIC | −1,329.290 | 27.030 | 101.780 | 335.430 | 4,951.780 | 230.750 | 366.594 |
| | | | BIC | −1,321.63 | 31.320 | 110.350 | 349.160 | 4,970.52 | 240.380 | 370.042 |
| 3 | ln (Average price per ad impression per day) | Day count | $R^2$ | .000 | .009 | .039 | .120 | .963 | .093 | .135 |
| | | | AIC | −472.400 | 53.900 | 149.500 | 396.700 | 2201.300 | 266.300 | 287.872 |
| | | | BIC | −464.750 | 57.830 | 156.650 | 406.960 | 2215.160 | 273.420 | 287.872 |
| 4 | ln (Average price per ad impression per day) | Day count ad inventory | $R^2$ | .000 | .052 | .122 | .246 | .989 | .176 | .169 |
| | | | AIC | −472.410 | 52.080 | 144.760 | 386.140 | 2149.510 | 258.930 | 281.427 |
| | | | BIC | −464.750 | 57.110 | 154.300 | 399.880 | 2168.050 | 268.570 | 285.693 |
| 5 | Average price per ad impression per day | Day count day count$^2$ ad inventory | $R^2$ | .000 | .060 | .131 | .269 | .998 | .196 | .187 |
| | | | AIC | -1,329.450 | 26.570 | 100.580 | 330.400 | 4,952.970 | 227.510 | 364.444 |
| | | | BIC | -1,319.230 | 31.970 | 111.380 | 347.230 | 4,976.400 | 239.530 | 368.654 |

Notes: This table reports estimates of cookie-specific regressions per reported model conditional on having at least ten observations per cookie. The resulting number of observations is 28,788. Ad inventory characteristics include media type, which captures the share of video ads over regular display ads per day; fold position, which captures the share of ads displayed above the fold per day; and the share of retargeted ad impressions per day. [a] The prediction quality measures $R^2$, MAE, and RMSE are obtained by training the model on the first 80% of consecutive observations per cookie and testing the model on the last 20% of consecutive observations per cookie. MAPE corresponds to the in-sample absolute difference between the observed lifetime value of the cookie and the predicted lifetime value of the cookie divided by the observed lifetime value of the cookie using the full data set. AIC: Akaike information criterion; BIC: Bayesian information criterion; MAE: mean absolute error; RMSE: root mean squared error; MAPE: mean absolute percentage error. [b] Model fit measures are calculated on the full data set.

# Web Appendix W5
# Simulation Study

In this Web Appendix, we outline how our simulation accommodates that (1) the predicted lifetime is longer than the observed lifetime, (2) cookies are not active every day, and (3) eliminates differences in the daily number of impressions per cookie across time.

**Consideration of Difference Between Predicted and Observed Lifetime of a Cookie**

Due to censoring, a cookie's actual lifetime (herein referred to as the uncensored lifetime of a cookie) can be larger than its observed (potentially censored) lifetime; thus, we must predict the price per ad impression for each day of the predicted residual lifetime beyond the observed lifetime. We use Cookie A to illustrate how we proceed if the observed lifetime is only 15 days but the uncensored lifetime is 22 days (as is the case for Cookie A in the main manuscript). In this case, the cookie generates a value of $.09 on day 1, $.10 on day 2, $.11 on day 3, and so forth, until $.23 on day 15. The observed lifetime value of this cookie (LVC) is thus the sum of the cookie values per day across the 15 days ($.09 + $.10 + $.11 +… + $.23 = $2.40). We then estimate our regression. The estimated time parameter outlines that the value per day increases by $.01. We use this information to predict the value per day for the remaining seven days (i.e., $.24 on day 16, $.25 on day 17, and so forth, to $.30 on day 22). The uncensored LVC is thus the sum of the cookie values per day across the 15 observed days (days 1–15: $2.40) and the 7 additional days (days 16–22: $1.89), thus $4.29.

**Consideration of the Difference Between the Number of Active and Observed Days**

So far, we have looked at cookies—more precisely, the users behind each cookie who were active each day (i.e., received an ad impression each day). Yet, cookies are also inactive on some days (i.e., they receive no ad impressions). We use Cookie A again to illustrate how we consider inactivity in our simulation study. Cookie A again has an observed lifetime of 15 days, and its uncensored lifetime is 22 days. Yet, Cookie A was inactive on day 7, that is, on one of the 15 days (= 1 / 15 = 6.66%). Thus, the share of observed active days per observed days of Cookie A is 93.33%.

We again estimate the regression and use the time parameter to predict the price of the ad impression on the inactive day (day 7) and the additional 7 days (days 16–22). The sum of those values is again $4.29. However, we now multiply the total value by the share of active days: 93.33% × $4.29 = $4.00. We also multiply the total value with the restriction ($2.89) with the share of active days, resulting in $2.70. As a result, LVC is reduced by $1.30, from $4.00 to $2.70 (−32%).

**Consideration of a Trend in the Number of Ad Impressions per Day**

So far, we have considered an equal number of ad impressions per day, namely one, such that the average price of an ad impression was equal to the revenue per day. However, revenue is the product of the number of ad impressions per day times the average daily price per ad impression; therefore, we must account for the difference between revenue and price per ad impression. Only changes in the price per ad impression reflect that an increase in information about the user (as made available by the cookie) reflects the cookie's value.

It is possible that the cookie changed activity over time; for example, it could become more active because the user behind the cookie learned how to use the internet better. These changes do not reflect the value of the cookie. Therefore, we need to isolate the changes in revenue per day from those changes due to changes in the average price, not those due to changes in the number of ad impressions.

We again use Cookie A to illustrate how we isolate those changes in our simulation study. Cookie A again has an observed lifetime of 15 days, its uncensored lifetime is 22 days, and it was inactive on day 7, that is, on one of the 15 days (= 1 / 15 = 6.66%). Thus, the share of observed active days at all observed days of Cookie A is 93.33%. Yet, daily ad impressions no longer remain constant but increase by 1 per day. So, we have 1 ad impression on day 1, 2 ad impressions on day 2, and so forth, until 15 ad impressions on day 15. The revenue of those 15 days is now $20.95.

Now, we run two regressions — one for the daily price (as we did before) and one for the daily ad impressions. The result for the "price regression" is (as before) price per impression and day = $.08 + $.1 × day. The result for the "quantity regression" is impressions per day = 0 + 1 × day. Thus, the time parameter in this regression represents the daily increase in ad impressions. We eliminate these changes in the number of ad impressions by replacing the daily number of impressions with the daily number of impressions predicted by the regression for the average value of the independent variable, which is 8.07 (= 0 + 1 × 8.07, as we ignore the inactive day 7). This value also equals the average daily number of impressions in the first 15 days (actually 14 days, as we ignore the inactive day 7), which is 8.07 (=113 / 14).

We then proceed as outlined previously. We calculate the revenue for the inactive day 7 ($1.21 = 8.07 × $0.15) and days 16–22 (e.g., for day 16: $1.94 = 8.07 × $.24). We multiply the sum of those values with the share of active days ($34.05 = 93.33% × $36.48), representing Cookie A's value without lifetime restrictions. We proceed similarly for the case of the lifetime restriction, yielding a value of $22.75 and thus a loss of $11.30 (−33.19%).

In the Supplemental Material to this paper, we provide a spreadsheet that outlines how we determine the economic consequences of cookie lifetime restrictions in our numerical example.

# Web Appendix W6
# Descriptive Regressions

TABLE W6.1

REGRESSION RESULTS OF COVARIATES PER COOKIE

ON CONSTANT AND TIME PARAMETER

| | | Dep. Variable: Constant | $p$-Value | Dep. Variable: Time Parameter[a] | $p$-Value |
|---|---|---|---|---|---|
| Country | National | .441 | .000*** | .212 | .837 |
| | International | .308 | .004** | −.171 | .868 |
| Device Type | Desktop | .168 | .000*** | .158 | .303 |
| | Mobile | .237 | .000*** | .325 | .160 |
| Operating System | Android | −.242 | .000*** | −.289 | .128 |
| | BlackBerry | −.327 | .000*** | −.557 | .004** |
| | Chrome | .120 | .000*** | −.429 | .050* |
| | iOS | −.071 | .000*** | −.394 | .032* |
| | Linux | −.156 | .000*** | −.004 | .986 |
| | PlayStation | .170 | .000*** | −.169 | .378 |
| | Windows | −.040 | .000*** | .073 | .137 |
| Browser | Android | .232 | .000*** | −.140 | .722 |
| | Chrome | .142 | .000*** | .012 | .974 |
| | Firefox | .116 | .003** | −.068 | .858 |
| | Internet Explorer | .029 | .320 | .137 | .720 |
| | iOS | −.120 | .006** | −.282 | .450 |
| | Opera | .189 | .000*** | .244 | .565 |
| | Safari | .284 | .000*** | −.005 | .990 |
| Constant | — | −.212 | .061 | .076 | .944 |
| Adj. $R^2$ | — | .036 | | .001 | |
| N | | 98,527 | | | |

Notes: This table shows the results of two regressions using the pooled data of samples 1 and 2.
$p$-values: * $p \leq .05$, ** $p \leq .01$, *** $p \leq .001$. The reference category for each categorical variable is "Unknown". [a] We multiplied the parameter estimates by 1,000 for ease of readability.

# Web Appendix W7
# Additional Results Sample 2

**Description of the Second Sample**

We repeat the analysis for our second sample, and Table W7.1 provides the descriptive statistics. The mean cookie age on our sampling day is 120 days (median: 28 days). We also find a mean observed cookie lifetime of 228 days (median: 94).

We again use a survival model to account for potential censoring of the observed cookie lifetime. Our second sample includes 5.399% of (potentially) left-censored cookies (i.e., we can only observe cookie death) and 6.313% of right-censored cookies (i.e., we can only observe cookie birth). 2.550% are both right- and left-censored (i.e., we can observe neither cookie birth nor cookie death). Thus, we have 14.262% of censored cookies in our second sample. We again fit a parametric Weibull and parametric Lognormal survival model to the data (see Table W7.2).

We select the Weibull model for further analysis because it fits the data better than the Lognormal model (Weibull: LL: −154,132.700, AIC: 308,269.500, BIC: 308,285.900 vs. Lognormal: LL: −154,630.200, AIC: 309,264.500, BIC: 309,280.900) (for validation of this approach, see Web Appendix W3). Again, we integrate the parametric Weibull distribution to find the residual cookie lifetime per cookie using the observed lifetime per cookie as the lower bound of the integral. The mean cookie lifetime adjusted for censoring is 298 days (median: 94 days).

The cookies of our second sample were active on average on 81 of the observed days (median: 11 days), yielding an average share of observed active days at the observed days of 55.300% (median: 54.000%). The cookies also differ concerning the observed number of ad impressions served. They reach an average number of 2,369 ad impressions (median: 47) and an average number of ad impressions per observed day of 9.064 (median: 2.172).

We calculate that cookie values per day in our second sample reach a mean value of €.005 (median: €.001). We also compute the average uncensored cookie value per day at €.006 (median: €.001). The mean value per 1,000 ad impressions paid by the purchasing advertiser is €.720 CPM (median: €.683 CPM).

Concerning the observed cookie lifetime value, we find a mean lifetime value of €1.569 (median: €.030). The mean predicted lifetime value is €1.791 (median: €.029). We calculate a MAPE of .083 (median: .003). The mean predicted residual lifetime value amounts to €.960 (median: .000.). The uncensored cookie lifetime value yields different results, with a mean cookie lifetime value of €2.799 (median: €.029).

**Results of Regression Analysis**

*Linear model without additional covariates.* We summarize the results of our regression analysis in Table W7.3. In model 1, we find a positive incremental effect for 5,890 cookies. They represent 13.266% of all cookies in our sample. They received 46.669% of all ad impressions, with an average number of ad impressions per cookie of 8,336. Their mean uncensored cookie lifetime is 629 days, and the average uncensored cookie's lifetime value is €10.123.

We find a significant negative incremental effect for 4,376 cookies, representing 9.856% of all cookies in our sample. They received 18.942% of all ad impressions, with an average number of ad impressions per cookie of 4,554. Their mean cookie lifetime is 586 days, and the average uncensored cookie lifetime value is €4.950.

For most cookies (34,134, i.e., 76.878%), the time parameter is insignificant, i.e., zero. Specifically, 10,196 cookies, or 22.964% of all cookies, have this "zero effect". Yet, these cookies received 34.389% of all ad impressions, with an average number of ad impressions per cookie of 1,060. Their mean uncensored cookie lifetime is 203 days, and the average uncensored cookie lifetime value is €1.205.

*Linear model with additional covariates.* In model 2, our preferred model specification (see Equation 1), we find a positive incremental effect for 6,106 cookies. They represent 13.752% of all cookies in our sample. They received 52.197% of all ad impressions, with an average number of ad impressions per cookie of 8,993. Their mean uncensored cookie lifetime is about 632 days, and the average uncensored cookie's lifetime value is €10.756.

We find a significant negative incremental effect for 4,097 cookies, representing 9.227% of all cookies in our sample. They received 16.245% of all ad impressions, with an average number of ad impressions per cookie of 4,172. Their mean uncensored cookie lifetime is 586 days, and the average uncensored cookie lifetime value is €4,761.

The time parameter is insignificant for most cookies (34,197 or 77.020%). Specifically, 10,132 cookies, or 22.820% of all cookies, have a zero effect in the sample. Yet, these cookies received 31.558% of all ad impressions, with an average number of ad impressions per cookie of 971. Their mean uncensored cookie lifetime is 203 days, and the average uncensored cookie lifetime value is €1.143.

Table W7.1

SAMPLE 2: SUMMARY STATISTICS PER COOKIE (N = 44,400)

| Category | Variable | Quantiles | | | | | Mean | SD |
|---|---|---|---|---|---|---|---|---|
| | | Min. | 25% | 50% | 75% | Max. | | |
| Lifetime Unit of Cookie | Observed Age of Cookie on Sampling Day (in days)[a] | 1 | 1 | 28 | 195 | 467 | 120 | 161 |
| | Observed (potentially censored) Lifetime of Cookie (in days)[a] | 1 | 1 | 94 | 428 | 867 | 228 | 270 |
| | Uncensored Lifetime of Cookie (in days)[a,b] | 1 | 1 | 94 | 440 | 1,359 | 298 | 406 |
| | Observed Number of Active Days | 1 | 1 | 11 | 90 | 839 | 81 | 144 |
| | Share of Observed Active Days per Observed Days | .001 | .143 | .540 | 1.000 | 1.000 | .553 | .393 |
| | Observed Number of Ad Impressions[c] | 1 | 5 | 47 | 971 | 504,999 | 2,369 | 9,977 |
| | Observed Number of Ad Impressions per Day | .003 | .896 | 2.172 | 8.373 | 1,017.200 | 9.064 | 24.717 |
| Value of Cookie per Lifetime Unit | Observed Value of Cookie per Day (in €) | .000 | .000 | .001 | .006 | .494 | .005 | .013 |
| | Uncensored Value of Cookie per Day (in €) | .000 | .000 | .001 | .006 | .674 | .006 | .016 |
| | Observed Value of Cookie per Ad Impression (in €, CPM) | .000 | .389 | .683 | .948 | 54.819 | .720 | .651 |
| Lifetime Value of Cookie | Observed (potentially censored) Lifetime Value of Cookie (in €) | .000 | .002 | .030 | .717 | 332.710 | 1.569 | 5.459 |
| | Predicted Censored Lifetime Value of Cookie for Observed Lifetime (in €) | .000 | .002 | .029 | .799 | 401.728 | 1.791 | 6.280 |
| | Mean Absolute Percentage Error (MAPE[d]) for Observed Lifetime | .000 | .000 | .003 | .102 | 6.278 | .083 | .193 |
| | Predicted Residual Lifetime Value of Cookie for Residual Lifetime (in €) | .000 | .000 | .000 | .000 | 167.159 | .960 | 5.242 |
| | Uncensored Lifetime Value of Cookie (in €)[e] | .000 | .002 | .029 | .892 | 453.524 | 2.799 | 11.066 |

Note: [a] Rounded to the next full day. [b] We use a Weibull model to determine the expected residual lifetime for 14,262% of the cookies with potentially censored cookie lifetime. The adjusted average predicted cookie lifetime of 298 days is 30.702% larger than the average observed cookie lifetime of 228 days in the data (i.e., sample 2). [c] MAPE corresponds to the in-sample absolute difference between the observed lifetime value of the cookie and the predicted lifetime value of the cookie divided by the observed lifetime value of the cookie. [e] We determine the uncensored cookie lifetime value using the regression outlined in Equation 1 (i.e., model 2 in Table 7).

TABLE W7.2
SAMPLES 2: SURVIVAL MODEL PARAMETERS AND FIT MEASURES

| Model | Shape Parameter [95%-CI] | SE | Scale Parameter [95%-CI] | SE | LL | AIC | BIC |
|---|---|---|---|---|---|---|---|
| | | | SAMPLE 2 (N = 44,400) | | | | |
| Weibull | .975 [.963; .986] | .006 | 466.783 [460.434; 473.219] | 3.262 | −154,132.700 | 308,269.500 | 308,285.900 |
| Lognormal | 5.641 [5.624; 5.658] | .009 | 1.402 [1.388; 1.416] | .007 | −154,630.200 | 309,264.500 | 309,280.900 |

Notes: To avoid that those cookies with very short lifetimes impact our results too strongly, we only consider cookies with an observed cookie lifetime of seven or more days to predict residual cookie lifetime. CI: confidence interval; SE: standard error; LL: loglikelihood value; AIC: Akaike information criterion; BIC: Bayesian information criterion

TABLE W7.3
REGRESSION RESULTS OF IMPACT OF TIME ON THE AVERAGE PRICE PER AD IMPRESSION PER DAY

|  | Significant Positive Incremental Effect | | Significant Negative Incremental Effect | | Nonsignificant Incremental Effect | |
| --- | --- | --- | --- | --- | --- | --- |
|  | Model 1 | Model 2 | Model 1 | Model 2 | Model 1 | Model 2 |
| Dependent Variable (in €; CPM) | Average Price per Ad Impression per Day | | | | | |
| Constant [95%- CI] | .487 [.325; .663] | .609 [.402; 845] | 1.130 [.923; 1.308] | 1.118 [.872; 1.335] | .000 [.000; .000] | .000 [.000; .000] |
| Time Parameter [95%- CI] | .001 [.000; .002] | .001 [.000; 002] | −.001 [−.002; .000] | −.001 [−.002; −.000] | .000 [.000; .000] | .000 [.000; .000] |
| Additional Covariates | N | Y | N | Y | N | Y |
| Number of Cookies (% of all cookies) | 5,890 (13.266%) | 6,106 (13.752%) | 4,376 (9.856%) | 4,097 (9.227%) | 34,134 (76.878%) | 34,197 (77.020%) |
| Total Number of Ad Impressions (% of all) | 49,095.327 (46.669%) | 54,910,981 (52.197%) | 19,926,405 (18.942%) | 17,089,452 (16.245%) | 36,177,071 (34.389%) | 33,198,370 (31.558%) |
| Ad Impressions per Cookie | 8,336 | 8,993 | 4,554 | 4,172 | 1,060 | 971 |
| Mean (Median) Uncensored Cookie Lifetime (in days) | 629 (482) | 632 (490) | 586 (448) | 586 (440) | 203 (12) | 203 (12) |
| Mean (Median) Uncensored Cookie Lifetime Value (in €) | 10.123 (2.927) | 10.756 (3.002) | 4.950 (1.081) | 4,761 (1.093) | 1.205 (.009) | 1.143 (.009) |
| Number of Cookies with Significant Zero Effect (% of all cookies) | — | — | — | — | 10,196 (22.964%) | 10,132 (22.820%) |

Notes: Unless otherwise noted, this table reports our sample's median estimates from 44,400 cookie-specific regressions. We consider the value increase (decrease) per day to be positive (negative) if the sign of the time parameter is positive (negative) and the value of the parameter is significant (at a 1% level). If the value is insignificant, then we conclude that there is no increase (decrease) in value over time. We apply a small Winsorization to accommodate outliers and replace the most extreme values with the 99% quantile of the respective parameter estimate. Model 1 only includes the time parameter (here: day count) as the independent variable. Model 2 includes the time parameter (here: day count) and additional covariates (i.e., ad inventory characteristics) as independent variables.

# Web Appendix References

Agencia Española de Protección de Datos (2020), "Guía Sobre el Uso de las Cookies," Spanish Data Protection Agency (July), https://www.aepd.es/sites/default/files/2020-07/guia-cookies.pdf.

Article 29 Data Protection Working Party (2010), "Opinion 2/2010 on Online Behavioural Advertising," Brussels: European Commission (June 22), https://ec.europa.eu/justice/article-29/documentation/opinion-recommendation/files/2010/wp171_en.pdf.

Article 29 Data Protection Working Party (2011), "Opinion 16/2011 on EASA/IAB Best Practice Recommendation on Online Behavioural Advertising," Brussels: European Commission (December 8), https://ec.europa.eu/justice/article-29/documentation/opinion-recommendation/files/2011/wp188_en.pdf.

Article 29 Data Protection Working Party (2013), "Working Document 02/2013 providing Guidance on Obtaining Consent for Cookies" (October 2), https://ec.europa.eu/justice/article-29/documentation/opinion-recommendation/files/2013/wp208_en.pdf.

Article 29 Data Protection Working Party (2015), "Cookie Sweep Combined Analysis Report" (February 3), https://ec.europa.eu/newsroom/article29/document.cfm?action=display&doc_id=56123.

Barker, Alex and Madhumita Murgia (2020), "Google to Phase Out Most Invasive Internet Tracking," *Financial Times* (January 14), https://www.ft.com/content/1f56591e-36e1-11ea-a6d3-9a26f8c3cba4.

Commission Nationale de l'Informatique et des Libertés (2020), "Délibération n° 2020-092 du 17 Septembre 2020 portant Adoption d'une Recommandation Proposant des Modalités Pratiques de mise en Conformité en cas de Recours aux ' Cookies et autres Traceurs'," French Data Protection Authority (September 17), https://www.cnil.fr/sites/default/files/atoms/files/recommandation-cookies-et-autres-traceurs.pdf.

Council of the European Union (2021), "Proposal for a Regulation of the European Parliament and of the Council Concerning the Respect for Private Life and the Protection of Personal Data in Electronic Communications and Repealing Directive 2002/58/EC (Regulation on Privacy and Electronic Communications)", (January 21), https://data.consilium.europa.eu/doc/document/ST-5008-2021-INIT/en/pdf.

Cook, Sam (2017), "How to Comply with Cookie Legislation and Respect Your Website Visitors' Privacy," Comparitech (February 24), https://www.comparitech.com/blog/vpn-privacy/how-to-comply-with-cookie-legislation.

Cookiebot (2022), "GDPR and Cookies | Compliant Cookie Use with Cookiebot CMP" (January 17), https://www.cookiebot.com/en/gdpr-cookies/.

Curac-Dahl, Peter and Marek Juszczyński (2021), "How a Cookie Audit Can Get You Up To Date in Today's Digital Privacy Landscape," Piwik Pro (April 15), https://piwik.pro/blog/cookie-audit-digital-privacy.

European Union (2017a), "Proposal for a Regulation of the European Parliament and the Council Concerning the Respect for Private Life and the Protection of Personal Data in Electronic Communications and Repealing Directive 2002/58/EC (Regulation on Privacy and Electronic Communications)," EUR-Lex (January 10), https://eur-lex.europa.eu/legal-content/EN/ALL/?uri=CELEX:52017PC0010.